\documentclass[12pt,a4paper,final]{iopart}

\usepackage{iopams}
\usepackage{graphicx}
\usepackage[breaklinks=true,colorlinks=true,linkcolor=blue,urlcolor=blue,citecolor=blue]{hyperref}
\graphicspath{{Figures/}}
\begin{document}

\title{Nernst effect in metals and superconductors: a review of concepts and experiments}

\author{Kamran Behnia and Herv\'e Aubin}
\address{Laboratoire de Physique et d'Etude de Mat\'eriaux\\
(Centre National de la Recherche Scientifique -Universit\'e Pierre et Marie Curie)\\
Ecole Sup\'erieure de Physique et de Chimie Industrielles\\
 75005 Paris, France}

\ead{kamran.behnia@espci.fr}

\begin{abstract}
 The Nernst effect is the transverse electric field produced by a longitudinal thermal gradient in presence of magnetic field. In the beginning of this century, Nernst experiments on cuprates were analyzed assuming that: i) The contribution of quasi-particles to the Nernst signal is negligible; and ii) Gaussian superconducting fluctuations cannot produce a Nernst signal well above the critical temperature. Both these assumptions were contradicted by subsequent experiments. This paper reviews  experiments documenting multiple sources of a Nernst signal, which, according to the Brigman relation, measures the flow of transverse entropy caused by a longitudinal particle flow. Along the lines of Landauer's approach to transport phenomena, the magnitude of the transverse magneto-thermoelectric response is linked to the quantum of thermoelectric conductance and a number of material-dependent length scales: the mean-free-path, the Fermi wavelength, the de Broglie thermal wavelength and the superconducting coherence length. Extremely mobile quasi-particles in dilute metals generate a widely-documented Nernst signal. Fluctuating Cooper pairs in the normal state of superconductors have been found to produce a detectable Nernst signal with an amplitude conform to the Gaussian theory, first conceived by Ussishkin, Sondhi and Huse. In addition to these microscopic sources, mobile Abrikosov vortices, mesoscopic objects carrying simultaneously entropy and magnetic flux, can produce a sizeable Nernst response. Finally, in  metals subject to a magnetic field strong enough to truncate the Fermi surface to a few Landau tubes, each exiting tube generates a peak in the Nernst response.  The survey of these well-established sources of the Nernst signal is a helpful guide to identify the origin of the Nernst signal in other controversial cases.

\end{abstract}

\vspace{2pc}
%\noindent{\it Keywords}:
%\submitto{\ROP}
%\maketitle

\section{Introduction}

Extending back to the mid-nineteenth century, the history of thermoelectricity can be seen as a continuous struggle to follow Kelvin's attempt to understand it as a quasi-thermodynamic phenomenon. The most important episode in this tale is Onsager's formulation of the reciprocal relations\cite{onsager1931}. In 1948, Callen\cite{callen1948} demonstrated that it provides a solid basis for the Kelvin relation between Seebeck and Peltier coefficients as well as the Bridgman relation between Nernst and Ettingshausen effects\cite{bridgman1924}.

Today, potential technological applications of materials with a sizeable thermoelectric figure of merit is the subject matter of a vast research activity\cite{goldsmid2010}. In this context,  the Nernst-Ettingshausen effect, the transverse thermoelectric response emerging in presence of magnetic field,  is much less explored than its longitudinal counterpart. Ironically, however, the record of the lowest temperature attained by thermoelectric cooling starting from room temperature is still held by an Ettingshausen cooler\cite{harman1964}, albeit in a magnetic field as strong as 11 T, which severely limits potential applications.

From a fundamental perspective, the Nernst effect attracted much attention following the discovery of a sizeable Nernst coefficient in a high-temperature cuprate superconductor at the turn of this century\cite{xu2000}. Numerous studies followed and led to a clarification of  how quasi-particles\cite{behnia2009} and Gaussian superconducting fluctuations\cite{pourret2009} can generate a Nernst signal. The purpose of the current paper is to give a picture of our current understanding of the Nernst effect in metals and superconductors. Since the publication of two previous review articles\cite{behnia2009,pourret2009} in 2009, experimental and theoretical studies have sharpened our overall picture of transverse thermoelectricity, which we wish to sketch in the present review.

The article begins by framing the significance of thermoelectric response in general and transverse thermoelectricity in particular following a picture first sketched by Callen, based on Onsager's reciprocity relations. This is helpful to identify the fundamental constants and the material-dependent length scales which set the magnitude of the transverse thermoelectric response, in a manner, which follows Landauer's approach to transport phenomena. Afterwards, we review experimental results and their interpretation in four different contexts where, quasi-particles,  short-lived Cooper pairs, superconducting  vortices and Landau tubes are believed to play a prominent role in producing a Nernst signal. We end this review by a brief discussion of other cases, not identified as belonging to any of these four categories.

\section{Callen's picture of thermoelectricity}
The application of Onsager's reciprocal relations to thermoelectric and thermomagnetic phenomena by Callen in 1948 put the Kelvin and Bridgman relations on a firm ground. Back in the nineteenth century, Lord Kelvin had linked the Seebeck and the Peltier effects. According to him, these two apparently distinct thermoelectric phenomena, the Seebeck, $S$, and the Peltier, $\Pi$, coefficients are simply proportional to each other. According to the Kelvin relation:
\begin{equation}
 \Pi= S T
\end{equation}

The Seebeck coefficient is the ratio of the electric field to the thermal gradient applied to generate it. On the other hand, the Peltier coefficient is equal to the ratio of heat current density to charge current density. Callen demonstrated that this link is a consequence of Onsager's reciprocity. His analysis led to an alternative definition of the Seebeck coefficient. Here are his own words: ``a current flowing in a conductor of a given temperature distribution carries with it an entropy per particle. This entropy flow is, of course, in addition to the entropy flow [driven by the temperature gradient]'' \cite{callen1948}. Therefore, it can be expressed as one of the two components of the entropy flow, $J_{\widetilde{S}}$:
\begin{equation}
\label{Eq:callen}
J_{\widetilde{S}}=S J_{e}- \kappa\frac{\nabla T}{T}
\end{equation}

Here, $J_{e}$ is the charge current density, $\kappa$ the thermal conductivity and $\nabla T$ the thermal gradient. As far as we know, this equation was first written down in ref.\cite{callen1948}.  According to it, in absence of thermal gradient,  the second term in the right hand vanishes and the Seebeck coefficient becomes simply the ratio of entropy flow to charge flow.

The counterpart of Kelvin relation for transverse thermoelectricity was first proposed in 1924 by Bridgman, who proposed a link between Nernst, $N$,  and  Ettingshausen, $\varepsilon$, coefficients through thermal conductivity, $\kappa$\cite{bridgman1924}:

\begin{equation}
\label{Eq:bridgman}
N = \varepsilon \kappa/ T 		
\end{equation}

The Nernst coefficient is the transverse electric field produced by a longitudinal thermal gradient and the Ettingshaussen coefficient is the transverse thermal gradient produced by a longitudinal charge current. Callen showed that the Bridgman relation can also be derived from Onsager reciprocal relations\cite{callen1948}. He argued that in presence of magnetic field, eight thermomagnetic coefficients can be derived from six independent Onsager coefficients, implying two unavoidable relations. One of these two is non-trivial and links \emph{isothermal} (and not adiabatic) Nernst, Ettingshaussen and thermal conductivity coefficients to each other. This was the same equation originally derived by Bridgman. In 1931, Sommerfeld and Frank had demonstrated its validity independently using a thermodynamic argument\cite{sommerfeld1931}.

Onsager's reciprocity leads to thermodynamic constraints in relations between what he called forces and fluxes. According to the Kelvin relation, measuring the ratio of the entropy flow to the particle flow is fundamentally equivalent to measuring the ratio of  the potential gradient (otherwise known as the electric field) to the thermal gradient. In other words, measuring ratio of two thermodynamic fluxes is expected to give an information identical to measuring the ratio of the thermodynamic forces which generate them. The ultimate reason behind the Bridgman relation is exactly the same. This can be seen by rewriting Eq.3 (assuming isotropic thermal conductivity):

\begin{equation}
\label{Eq:bridgman2}
\frac{E_{y}}{\nabla_{x}T} =\frac{\nabla_{y}T}{J^{e}_{x}} \frac{J^{Q}_{y}}{\nabla_{y}T}/ T = \frac{J^{\widetilde{S}}_{y}}{J^{e}_{x}}		
\end{equation}

In other words, the Nernst coefficient is a measure of transverse entropy flow caused by longitudinal particle flow.  As in the case of the Seebeck coefficient, this alternative definition provided by the application of Onsager reciprocity is strictly identical to the ratio of the transverse electric field to  the longitudinal thermal gradient invoked by the more common definition. This picture is particularly useful in setting a frame for the magnitude of the thermoelectric response linked to fundamental constants and material dependent parameters.

\section{Landauer formalism and thermoelectricity}

Let us now focus  on the particular case of mobile fermionic quasi-particles. The two equations linking the three conductivity tensors, $\overline{\sigma}$(electric), $\overline{\kappa}$ (thermal) and $\overline{\alpha}$ (thermoelectric), two flux densities, $\mathbf{J_{e}}$(charge current), and $\mathbf{J_{q}}$(heat current), and two driving vectors, $\mathbf{E}$(electric field) and  $\mathbf{\nabla T}$(thermal gradient) are often stated as\cite{ziman1964}:

\begin{equation}
\label{Eq:sigma_alpha}
    \mathbf{J_{e}}=\overline{\sigma}. \mathbf{E} - \overline{\alpha}.\mathbf{\nabla T}
\end{equation}
\begin{equation}
\label{Eq:alpha_kappa}
    \mathbf{J_{q}}= T \overline{\alpha} . \mathbf{E} - \overline{\kappa}.\mathbf{\nabla T}
\end{equation}

Each diagonal and off-diagonal component of these three tensors is set each by one of the six independent Onsager coefficients. A standard treatment using Boltzmann equation considering the electric field and thermal gradient as perturbations to a Fermi-Dirac distribution leads to two independent relations between the three tensors\cite{ziman1964}. One of these two, linking $\sigma$ and $\alpha$ is a cornerstone of fermionic thermoelectricity and is called the Mott relation\cite{mott1936}:

 \begin{equation}
\label{Mott}
\alpha=\frac{\pi^{2}}{3}\frac{k_{B}}{e}k_{B}T\frac{\partial\sigma}{\partial\epsilon}|_{\epsilon=\epsilon_{F}}
\end{equation}

In the Drude-Boltmann picture, the simplest expression for charge conductivity is:

\begin{equation}
\label{drude}
   \sigma_{xx} = ne\mu
\end{equation}

Here $n$ is the concentration of the carriers and $\mu$ their mobility. In the past few decades, a conceptual revolution initiated by Landauer\cite{landauer1957} led to consider conduction as transmission\cite{imry1999}. In this picture, an alternative transcription of this equation in two dimensions is :

\begin{equation}
\label{landauer}
   \sigma_{xx}^{2D} =2\pi \frac{e^{2}}{h}\frac{\ell}{\lambda_{F}}
\end{equation}

Eq.\ref{drude} and Eq.\ref{landauer} are strictly equivalent since one can write mobility as $\mu=\frac{e}{\hbar}\frac{\ell}{k_{F}}$. However Eq.\ref{landauer} transparently states that electric conductivity is  the capacity to transmit a quantum of electric conductance amplified or hindered by two material-dependent length scales: The distance electrons can go before being scattered ($\ell$ is the man-free-path) and their Fermi wave-length, $\lambda_{F}$. In other dimensions, the numerical prefactor and the exponent for $\lambda_{F}$ are different, but the picture is fundamentally similar.

The  thermoelectric counterpart of Eq.\ref{landauer} was  first discussed in the context of thermoelectricity of one-dimensional Quantum Point Contacts by van Houten \emph{el al.}\cite{houten1992}, who argued that the Mott relation should hold between $\sigma$ and $\alpha$ in one dimension. In two dimensions, using the Mott relation, one finds (See ref. \cite{behnia2015} for a more detailed discussion of the expression in different dimensions):
\begin{equation}
\label{landauer:TE}
   \alpha_{xx}^{2D} = \frac{2\pi^{2}}{3}\frac{k_{B}e}{h}\frac{\lambda_{F}\ell}{\Lambda^{2}}
\end{equation}

The new length scale introduced is the de Broglie thermal wavelength:

\begin{equation}
\Lambda^{2}= \frac{h^{2}}{2\pi m^{*} k_{B}T }
\end{equation}

In other words, the magnitude of  $\alpha_{xx}$ is set by the quantum of thermoelectric conductance ($\frac{k_{B}e}{h}$)and three material-dependent properties, which are $\lambda_{F}$, $\ell$ and $\Lambda$. In absence of a well-established terminology, we will call $\alpha_{xx}$ ($\alpha_{xy}$) the longitudinal (the transverse) thermoelectric coefficient. Usually they are not directly measured by experiment. Much more straightforward is to probe how the system responds to a quantified thermal gradient, which leads to a direct experimental measurement of the Seebeck and Nernst coefficients. The expression for the Seebeck coefficient simply becomes:

\begin{equation}
\label{SEEBECK}
   S_{xx} = \frac{\alpha_{xx}}{\sigma_{xx}}=\frac{\pi}{3}\frac{k_{B}}{e} (\frac{\lambda_{F}}{\Lambda})^2
\end{equation}

The Seebeck coefficient is a transport property. Nevertheless, in the simplest conceivable Fermi liquid,  its magnitude does not depend on the distance between two scattering events. This is a consequence of Callen's insight that the Seebeck coefficient is a ratio of two Onsager flows, and therefore, the ratio of the entropy of each electron to its electric charge. Their entropy is set by how much the wavelength of a traveling electron can deviate from the Fermi wavelength in the statistical distribution. We will use this picture in the next section to find what is the expected magnitude of $\alpha_{xy}$.

\section{Magnitude of transverse thermoelectric response}

\subsection{Quasi-particles} An intuitive window to the magnitude of $\alpha_{xy}$, the transverse thermoelectric response,  can be obtained by considering two more familiar cases. The longitudinal thermoelectric response (the Seebeck effect) and the transverse charge conductivity (the Hall effect) of the Fermi liquids.

\begin{figure}
\center
\begin{center}
\resizebox{!}{0.5\textwidth}{\includegraphics{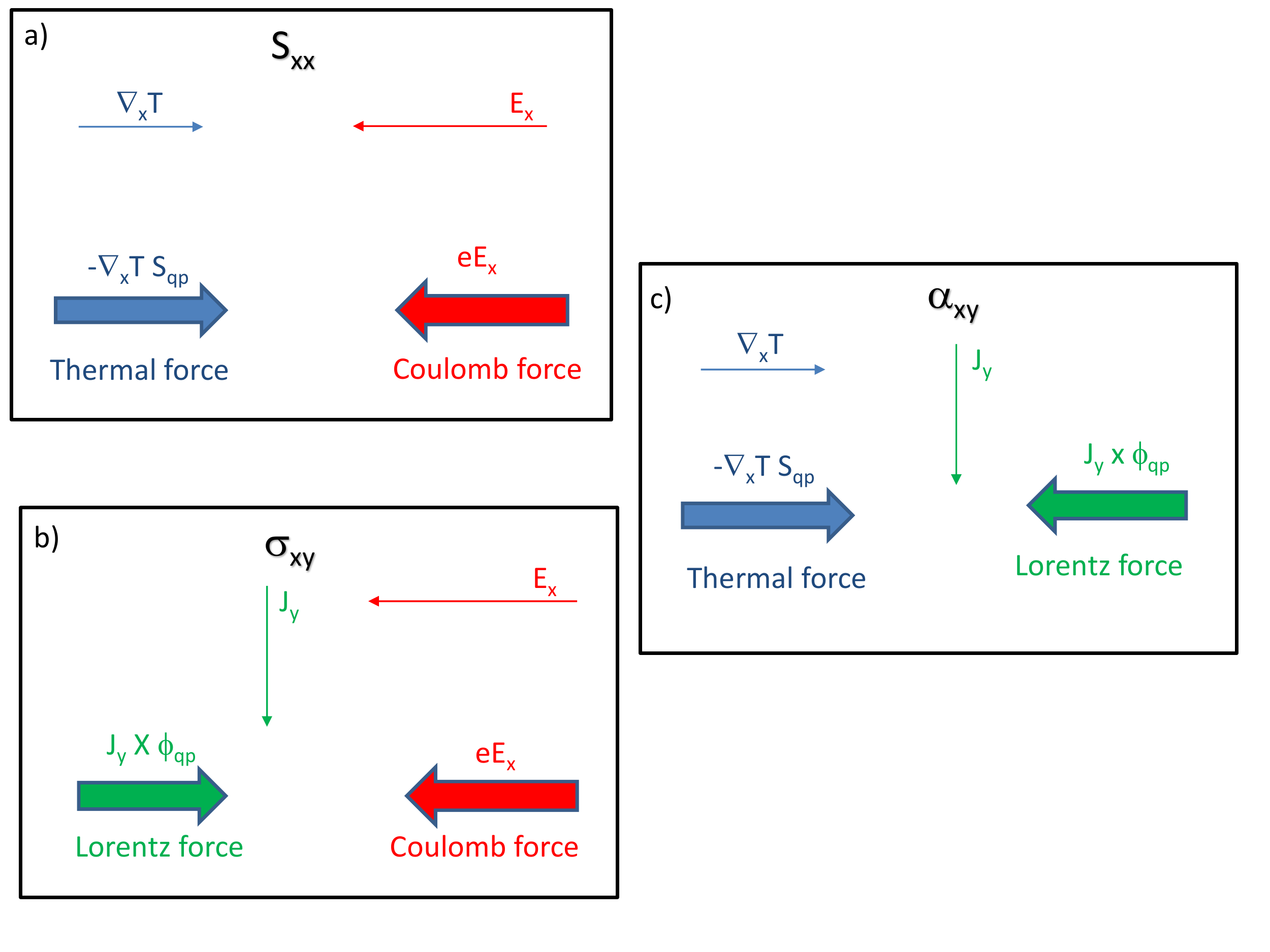}}
\end{center}
\caption{Seebeck, Hall and Nernst effects resulting from compensating forces: a) Carriers of electric charge, $e$  and entropy, S$_{qp}$ suffer a thermal force (driven by thermal gradient) and a Coulomb force (driven by electric charge). The magnitude of the Seebeck coefficient is set by the cancellation of these two forces.  b) Carriers of electric charge and magnetic flux, $\phi_{qp}$  suffer a Coulomb force and a Lorenz force (driven by a transverse charge current). The magnitude of the Hall coefficient is set by the cancellation of these two forces. c) Carriers of magnetic flux and entropy  suffer a thermal force and a Lorenz force (driven by a transverse charge current). The magnitude of the transverse thermoelectric coefficient, $\alpha_{xy}$ is set by their cancellation.}
\end{figure}

Longitudinal thermoelectricity arises when charged particles with a finite amount of entropy are simultaneously subject to an electric force and a thermal force (See Fig. 1a). In a solid with mobile electrons, the first is proportional to the electric field (and the electric charge) and the second is equal to the thermal gradient (and the entropy associated with each electron). In equilibrium, these two forces cancel each other. Therefore, the Seebeck coefficient, the ratio of the electric field to the thermal gradient is also the ratio of quasi-particle entropy, $\check{S}_{qp}$ to its charge, $e$:
\begin{equation}
\label{Eq:Sxx1}
S_{xx}= \frac{\check{S}_{qp}}{e}
\end{equation}

Now, the common expression for the Seebeck coefficient of a Fermi liquid is:
\begin{equation}
\label{Eq:Sxx2}
 S_{xx}=\frac{\pi^{2}}{3}\frac{k_{B}}{e}\frac{k_{B}T}{\epsilon_{F}}.
 \end{equation}

Are these two expressions equivalent?  The answer is affirmative. In a Fermi-Dirac distribution entropy available to transport is restricted to a thermal window, k$_{B}$T in the immediate vicinity of the Fermi energy, $\epsilon_{F}$. This leads to an expression of $\check{S}_{qp}$ in terms of two length scales\cite{behnia2015}:

\begin{equation}
\label{Eq:Sqp}
\check{S}_{qp}\simeq k_{B}(\frac{\lambda_F}{\Lambda})^{2}
\end{equation}

This is the origin of the equivalency between  Eq.\ref{Eq:Sxx1} and Eq.\ref{SEEBECK} or Eq.\ref{Eq:Sxx2}.

Let us now consider the Hall effect (Fig. 1b). Quasi-particles carrying both an electric charge, $e$ and a magnetic flux, $\phi_{qp}$, are subject to a coulomb force, eE$_{x}$, caused by a longitudinal electric field, E$_{x}$ and a Lorentz force, J$_{y}\times \phi_{qp}$ generated by a transverse charge current J$_{y}$. The compensation between these two forces set the magnitude of the Hall coefficient:

\begin{equation}
\label{Eq:sigmaxy}
\frac{J_{y}}{E_{x}}=\sigma_{xy}=\frac{e}{\phi_{qp}}
\end{equation}

What is the magnitude of $\phi_{qp}$? To answer this question, consider the Drude-Boltzmann picture of Hall conductivity in a simple two-dimensional circular Fermi surface:

\begin{equation}
\label{Eq:sigmaxy}
\sigma_{xy}|_{2D}= n e \mu \frac{\mu B }{1+ \mu^{2}B^{2}}
\end{equation}

This expression can be rewritten in terms of $\frac{ e^{2}}{h}$, material-dependent length scales and the magnetic length, $\ell_{B}=(\frac{\hbar}{e B})^{1/2} $:

\begin{equation}
\label{Eq:sigmaxy_bis}
\sigma_{xy}|_{2D}= \frac{e^{2}}{h} \frac{\ell^{2}\ell_{B}^{2}}{\ell_{B}^{4}+\ell^{2}k_{F}^{2}}
\end{equation}

In the weak-field limit ($\mu B \ll 1$), as Ong showed in a paper devoted to the geometric interpretation of the Hall conductivity\cite{ong1991}, it becomes simply:

\begin{equation}
\label{Eq:sigmaxyOng}
\sigma_{xy}|_{2D}= \frac{ e^{2}}{h}(\frac{\ell}{\ell_{B}})^2
\end{equation}

Combining Eq. \ref{Eq:sigmaxy} and Eq.\ref{Eq:sigmaxyOng}, one finds that the magnetic flux associated with each quasi-particle is:
\begin{equation}
\label{Eq:phiqp}
\phi_{qp}= \frac{h}{e}(\frac{\ell_B}{\ell})^2
\end{equation}

\begin{figure}
\center
\begin{center}
\resizebox{!}{0.5\textwidth}{\includegraphics{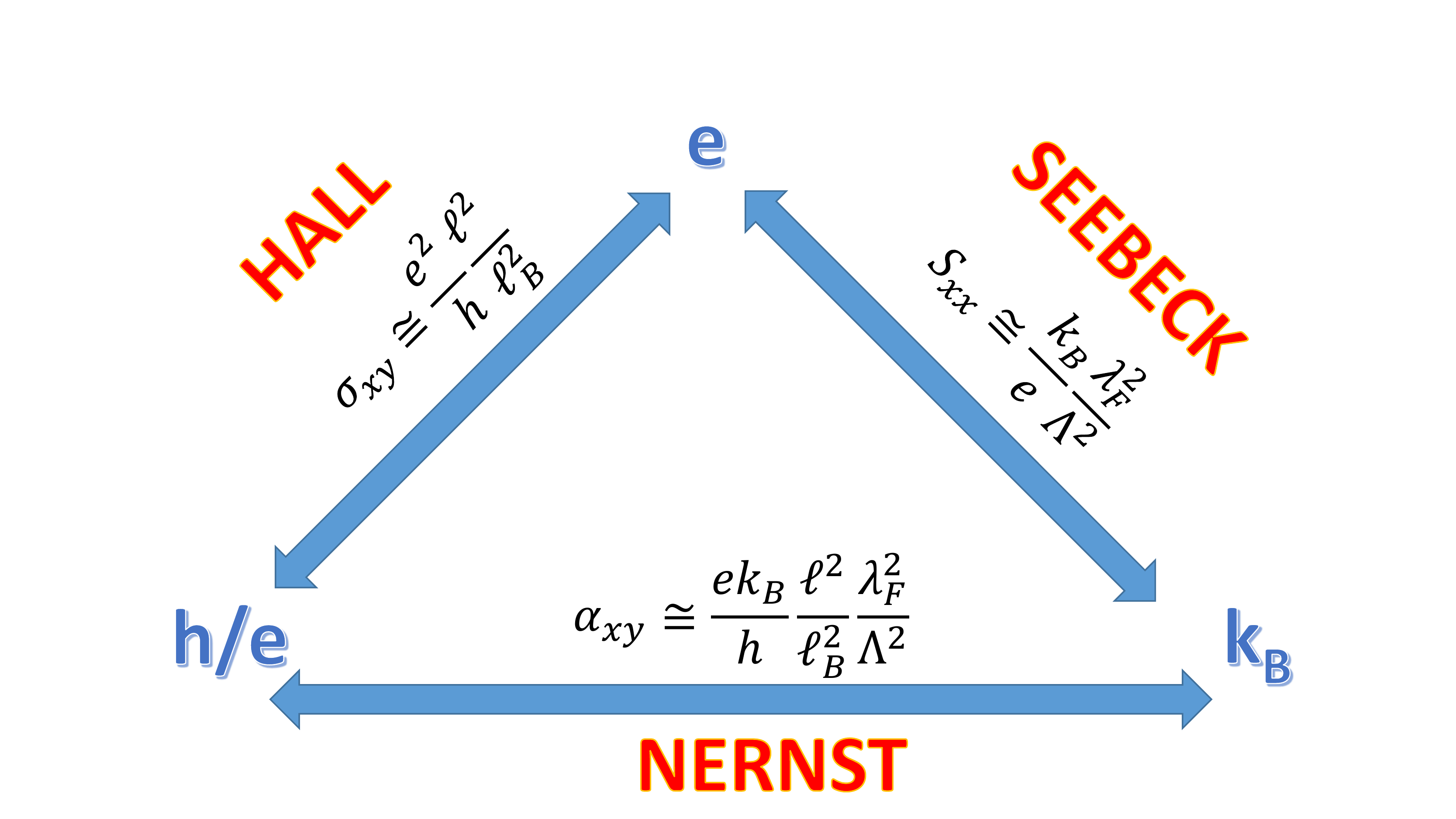}}
\end{center}
\caption{The link between three transport properties in a Fermi liquid. The Hall effect couples  electric charge and the magnetic flux.  The Seebeck effect couples charge to entropy. The Nernst effect (or more precisely the transverse thermoelectric coefficient, $\alpha_{xy}$) is a direct coupling between magnetic flux and entropy. The expressions given are those expected for a 2D circular Fermi liquid (see text).}
\end{figure}

With these two cases in mind, let us turn our attention to transverse thermoelectricity. Consider a system in which each carrier is a package containing both magnetic flux and entropy (Fig. 1c). Such carriers can suffer a thermal force proportional to their entropy (and the thermal gradient) and a Lorentz force proportional to their magnetic flux (and the transverse charge current). In equilibrium these two forces cancel out. Thus, the ratio of the transverse charge current to the longitudinal thermal gradient which generate it (the definition of $\alpha_{xy}$) is equal to the ratio of entropy to magnetic flux of traveling quasi-particles. In other words:
\begin{equation}
   \frac{J^{e}_{y}}{\nabla_{x}T}=\alpha_{xy}=\frac{\check{S}_{qp}}{\phi_{qp}}
\end{equation}

Here, $\check{S}_{qp}$ is the entropy of each carrier and $\phi_{qp}$ its magnetic flux.  Using the expressions for $\check{S}_{qp}$ (Eq. \ref{Eq:Sqp}) and $\phi_{qp}$ (Eq. \ref{Eq:phiqp}), one can derive the following expression for transverse thermoelectric conductivity:

\begin{equation}
\label{Eq:alphaqp}
\alpha_{xy}^{qp}|_{2D} \simeq \frac{ k_{B}e}{h}(\frac{\ell}{\ell_{B}})^2 (\frac{\lambda_{F}}{\Lambda})^2
\end{equation}

According to this equation,, the low-field transverse thermoelectric response is set by it natural units, $\frac{ek_{B}}{h}$, and three material-dependent length scales: the mean-free-path, the Fermi wavelength and the de Broglie thermal length.  Fig. 2 is summary of this very simple picture of quasi-particle transport. The transverse thermoelectric response is amplified by a long mean-free-path as well as the degeneracy intensity of the fermionic system.

\subsection{Short-lived Cooper pairs} Let us consider other sources for a Nernst signal in presence of a superconducting ground state. The gaussian fluctuations of the superconducting order parameter are one such source. Because of these fluctuations, short-lived Cooper pairs appear above the critical temperature and in the normal state. The expected  magnitude of their transverse thermoelectric response was first calculated by Usshishkin, Sondhi and Huse\cite{ussishkin2002}:

\begin{equation}
\alpha_{xy}^{scf}|_{2D}= \frac{1}{3}\frac{ k_{B}e}{h}(\frac{\xi}{\ell_{B}})^2
\end{equation}

Here $\xi$ is the superconducting coherence length. This expression can be intuitively understood by considering that each short-lived Cooper pair has an entropy of the order of k$_{B}$ and carries a magnetic flux equal to:
\begin{equation}
\label{Eq:phicp}
\phi_{cp}= \frac{h}{2e}(\frac{\ell_B}{\xi})^2
\end{equation}

Let us note that  the Cooper pairs, in contrast with quasi-particles, present a Nernst response which does not depend on the mean-free-path. This is also true of their paraconductivity, expressed as :
\begin{equation}
\sigma^{cp}_{xx}=\frac{n_s e^2\tau_{GL}}{m^*}
\end{equation}

Here, n$_{s}$ the concentration of Cooper pairs and $\tau_{GL}$ is the Ginzburg-Landau time scale. Unlike fermionic quasi-particles, the Landauer transmission factor does not depend on scattering time, but only on the pair lifetime, which diverges at the critical temperature. This time scale sets the temperature dependence of both the Nernst response and the paraconductivity.

\begin{table}
\centering
\begin{tabular}{c c c c }
  \hline
  \hline
carriers  & entropy  & flux  & $\alpha_{xy}|_{2D}$   \\
   \hline
 Quasi-particles &$\frac{\pi}{3}k_{B}(\frac{\lambda_F}{\Lambda})^2 $ & $\sim\frac{h}{e}(\frac{\ell_B}{\ell})^2$ & $\frac{k_{B}e}{h}(\frac{\ell}{\ell_{B}})^2 (\frac{\lambda_{F}}{\Lambda})^2 $  \\
 \hline
 Short-lived Cooper pairs & $k_{B}$ & $\frac{h}{2e}(\frac{\ell_{B}}{\xi})^2$ & $\sim\frac{ k_{B}e}{h}(\frac{\xi}{\ell_{B}})^2 $\\
   \hline
Superconducting vortices& $S_{vort}$ & $\frac{h}{2e}$ &$\sim\frac{e S_{vort}}{h }$ \\
  \hline
\end{tabular}
\caption{Entropy, magnetic flux and transverse thermoelectric response in the case of different types of carriers.}
\end{table}
\subsection{Superconducting vortices} Historically, a picture of transverse thermoelectricity similar to the one sketched above first, was used for superconducting vortices (See for example ref.\cite{huebener1979}).

In their case, one can also write:
\begin{equation}
\alpha_{xy}^{vort}|_{2D}= \frac{S_{vort}}{\phi_{0}}
\end{equation}

The superconducting vortex is a mesoscopic object carrying a magnetic flux equal to the quantum of magnetic flux $\phi_{0}=\frac{h}{2e}$ and an excess entropy of $S_{vort}$ in its core. The latter is set by the entropy balance between the normal and superconducting states.

The vortex Nernst signal has been often analyzed in the following way. In a Nernst experiment, the thermal force on vortices $\mathbf{f}=S_{vort}(-\mathbf{\nabla T})$ is balanced by a frictional force $\mathbf{f}_f=\eta \mathbf{v}$, where $\eta$ quantifies the viscosity of the vortices. The latter also sets the \emph{flux-flow} resistivity, $\rho=B\phi_0/\eta$, where $\phi_0=h/2e$ is the flux quantum associated with each vortex.

Using the Nernst data and the resistivity data one can eliminate the unknown $\eta$. Therefore, one can also write for vortices:
\begin{equation}
\label{NernstVortex}
\frac{N}{\rho}= \alpha_{xy}= \frac{S_{vort}}{\phi_0}
\end{equation}

Table 1 summarizes the magnetic flux, the entropy and the expected transverse thermoelectric response for these three sources of a Nernst signal.

\section{Review of experiments I: Quasi-particles}
Retrospectively, it is not surprising that Nernst and Ettingshausen discovered the effect that bears their names in bismuth. Both the mean-free-path and the Fermi wavelength are exceptionally long in this semimetal and according to Eq.\ref{Eq:alphaqp}, this would generate a very large transverse thermoelectric response.

Experiments have quantified the Nernst signal ($N=\frac{-E_{y}}{\nabla_{x}T}$), which in the low-field limit is related to $\alpha_{xy}$ though:

\begin{equation}
\label{Nalphaxy}
    N= \frac{\alpha_{xy}}{\sigma_{xx}}
\end{equation}

Now, In the light of this equation, as well as Eq.\ref{Eq:alphaqp} and Eq.\ref{landauer}, one can see that the picture sketched in the previous section is equivalent to the following expression for the Nernst coefficient in a metal\cite{behnia2009}:
\begin{equation}
\label{nuqp}
    \nu=\frac{N}{B}\sim\frac{\pi^{2}}{3}\frac{k_{B}}{e}\frac{k_{B}T}{\epsilon_F} \mu
\end{equation}

In other words, the Nernst coefficient of a given material hosting an electron fluid is set by two properties of the fluid: its Fermi energy, $\epsilon_{F}$, and its mobility, $\mu$. The higher this ratio, the larger is the expected Nernst response at low temperature. Note that Eq. \ref{Nalphaxy} is valid in both two and three dimensions, because the dimension-related length scale in $\alpha_{xy}$ and $\sigma_{xx}$ cancel out.
\begin{figure}
\begin{center}
\resizebox{!}{0.7\textwidth}{\includegraphics{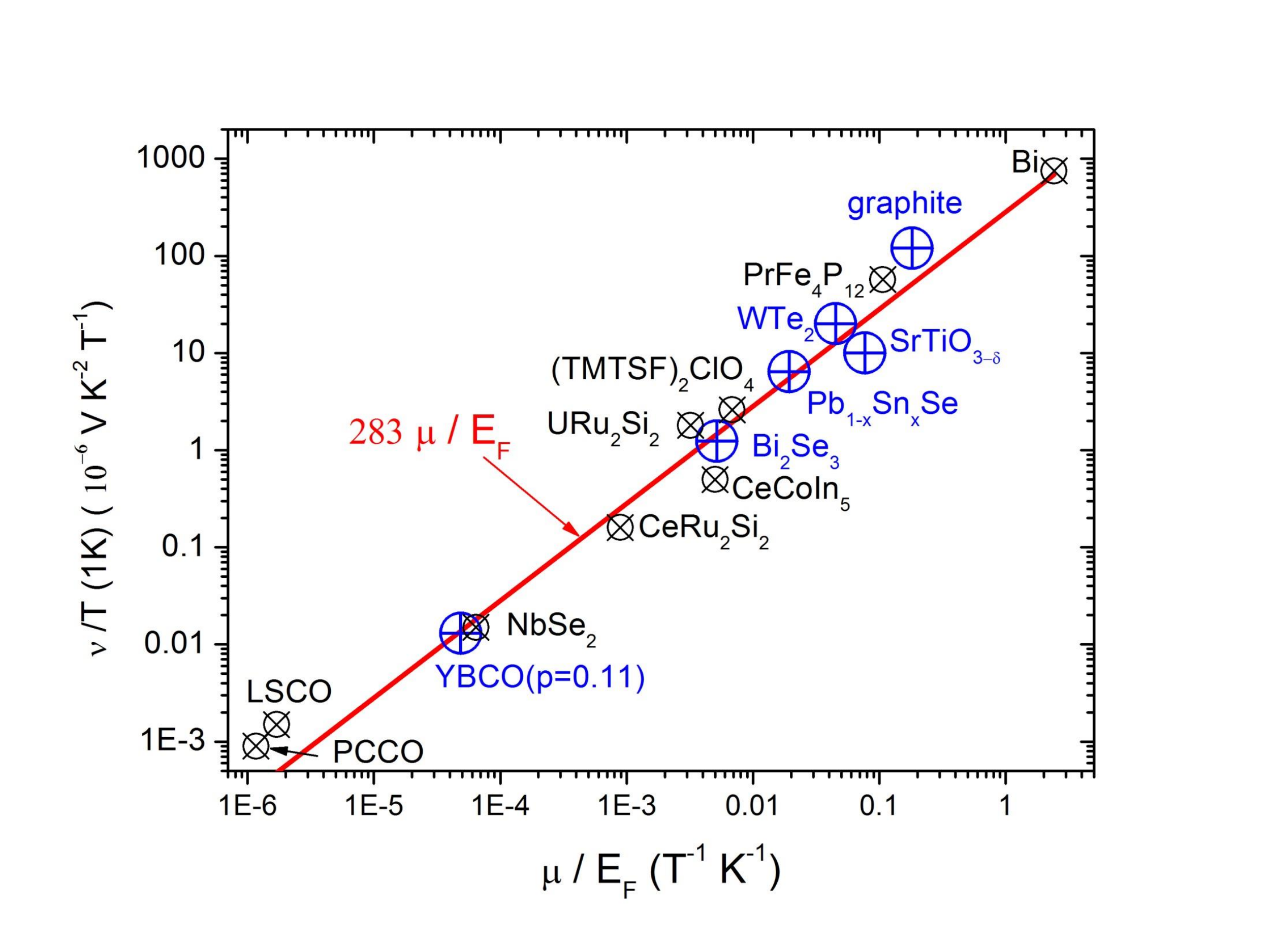}}
\caption{\label{Fig_Nernst_EF}The magnitude of the low temperature Nernst coefficient divided by temperature in a variety of metals plotted vs. the ratio of mobility to Fermi energy. The figure first appeared in a previous review article\cite{behnia2009}. Six blue data points represent those reported since its publication in 2009. The list two semimetals (graphite and WTe$_2$), three doped semiconductors (Bi$_{2}$Se$_{3}$ (at a carrier density of 10$^{17}$cm$^{-3}$) ,SrTiO$_{3}$ (at a carrier density of 5.5$\times$ 10$^{17}$cm$^{-3}$) and Pb$_{1-x}$Sn$_{x}$Se) and the high-T$_{c}$ cuprate, YBCO, at a single doping level of p=0.11, where small Fermi surface pockets have been resolved by experiment. }
\end{center}
\end{figure}

The large Nernst coefficient found in numerous heavy-fermion metals observed in the beginning of the twenty-first century was unexpected at first. However, given the low Fermi energy of these systems, several orders of magnitude lower than in high-density uncorrelated metals, their large Nernst response is unsurprising. Instead of enhancing the effective mass of electrons, one can pull down the Fermi energy by reducing carrier concentration. The lower the carrier concentration, the lower will be the size of the Fermi energy. Dilute metals present a large Nernst coefficient because of their low Fermi energy. On top of this, since small crystal imperfections cannot scatter extended objects, the long Fermi wavelength leads to a large mobility, providing a second source of enhancement for the Nernst response.

How quantitatively successful this approach is? It happens that as in the case of the Seebeck coefficient\cite{behnia2004}, an approach as simplistic as this gives a fair account of the magnitude of the low-temperature Nernst response scattered over many orders of magnitude in different metals. In a previous review article\cite{behnia2009}, the available data on a variety of metals was compared with the predictions of this expression and a satisfactory agreement was found.  Since then, the Nernst coefficient has been  measured in several other low-density systems.

Graphite is a semimetal with a small concentration (n$_{e}$=n$_{h}= 4\times 10^{18} cm^{-3}$) of high-mobility carriers of both signs\cite{brandt1988}. A large Nernst coefficient and its quantum oscillations at moderate magnetic field were resolved in graphite in 2010\cite{zhu2010}. The magnitude of the  Nernst coefficient resolved in the low-field limit was close to what is expected according to Eq.\ref{nuqp}. Another semimetal, WTe$_{2}$, attracted much attention following the discovery of its large unsaturated magnetoresistance persisting in a field as large as 60 T\cite{ali2014}. A recent study has explored its low-temperature thermoelectric properties, its Fermi surface structure\cite{zhu2015} and  its carrier density ($7\times10^{19}cm^{-3}$). The Nernst response was found to be linear in magnetic field in an extended window and its magnitude in rather good agreement with the expectations of Eq. \ref{nuqp}, given the mobility and the Fermi energy of the system.

In addition to semimetals, another group of dilute metals has been subject to recent thermoelectric studies. These are semiconductors sufficiently doped to be on the metallic side of the metal-insulator transition\cite{mott1990}. The effective Bohr radius can become long as a consequence of a large dielectric constant, a low effective mass or a combination of both. In such a case, the doped semiconductor on the metallic side of the metal-insulator transition hosts high-mobility carriers \cite{behnia2015b}, which give rise to quantum oscillations detectable at remarkably low magnetic fields. Because of the low Fermi energy and high electronic mobility, one expects to see a large Nernst response in such systems and indeed this has been observed in two very different metallic semiconductors\cite{fauque2013,lin2013}.

The bulk semiconductor, Bi$_{2}$S$e_{3}$, with a direct band gap of 0.2 eV has been known for decades as a sibling of Bi$_{2}$Te$_{3}$, the best thermoelectric material at room temperatures\cite{goldsmid2010}. In 2009, it was identified as a topological insulator in its stoichiometric composition\cite{zhang2009}. However, as a consequence of uncontrolled doping, the single crystals currently available are metals with a Fermi surface, hosting extremely mobile carriers\cite{butch2010}. The Nernst coefficient was recently measured in two Bi$_{2}$S$e_{3}$ single crystals with carrier concentrations ranging from $10^{17}$ to $10^{19}$ cm$^{-3}$. The Fermi energy substantially changes with carrier density and the magnitude of the measured signal has been found to be close to what is expected according to Eq.\ref{nuqp} in both samples\cite{fauque2013}.

The wide gap semiconductor, SrTiO$_{3}$ displays a metallic behavior for carrier concentrations exceeding 10$^{17}cm^{-3}$\cite{spinelli2010}. The resistivity of this dilute metal changes by three orders of magnitude upon cooling from room temperature to helium temperatures. There are three well-known routes for n-doping this semiconductor. One can either substitute strontium with lanthanum, titanium with niobium or simply remove oxygen. Oxygen reduction is particularly attractive as a simple way to finely tune the carrier concentration\cite{spinelli2010,lin2013}.  At a carrier concentration of $5.5\times10^{17}cm^{-3}$, the magnitude of the measured Nernst coefficient in this system is in very good agreement with the expected value given the mobility and the Fermi energy of the system.

The Nernst effect in  Pb$_{1-x}$Sn$_{x}$Se (x=0.23), a member of the IV-VI family of narrow-gap semiconductors was recently measured in 2013\cite{liang2013}. The n-doped samples had a carrier density of $3.46\times10^{17}cm^{-3}$ and a Hall mobility of  114,000 cm$^{-2}$V$^{-1}$s$^{-1}$.  The Fermi energy of the system was estimated to be 51 meV. At T=4.7 K, the study found a Hall  mobility of $\mu$= 114,000 cm$^{2}$ V$^{−1}$ s$^{−1}$ and a low-field Nernst coefficient  of $\nu \sim 40 \mu$ VK$^{-1}T^{-1}$. As seen in Fig. \ref{Fig_Nernst_EF}, these numbers put this system in company of the others.
\begin{figure}
\begin{center}
{\includegraphics[width=9cm,keepaspectratio]{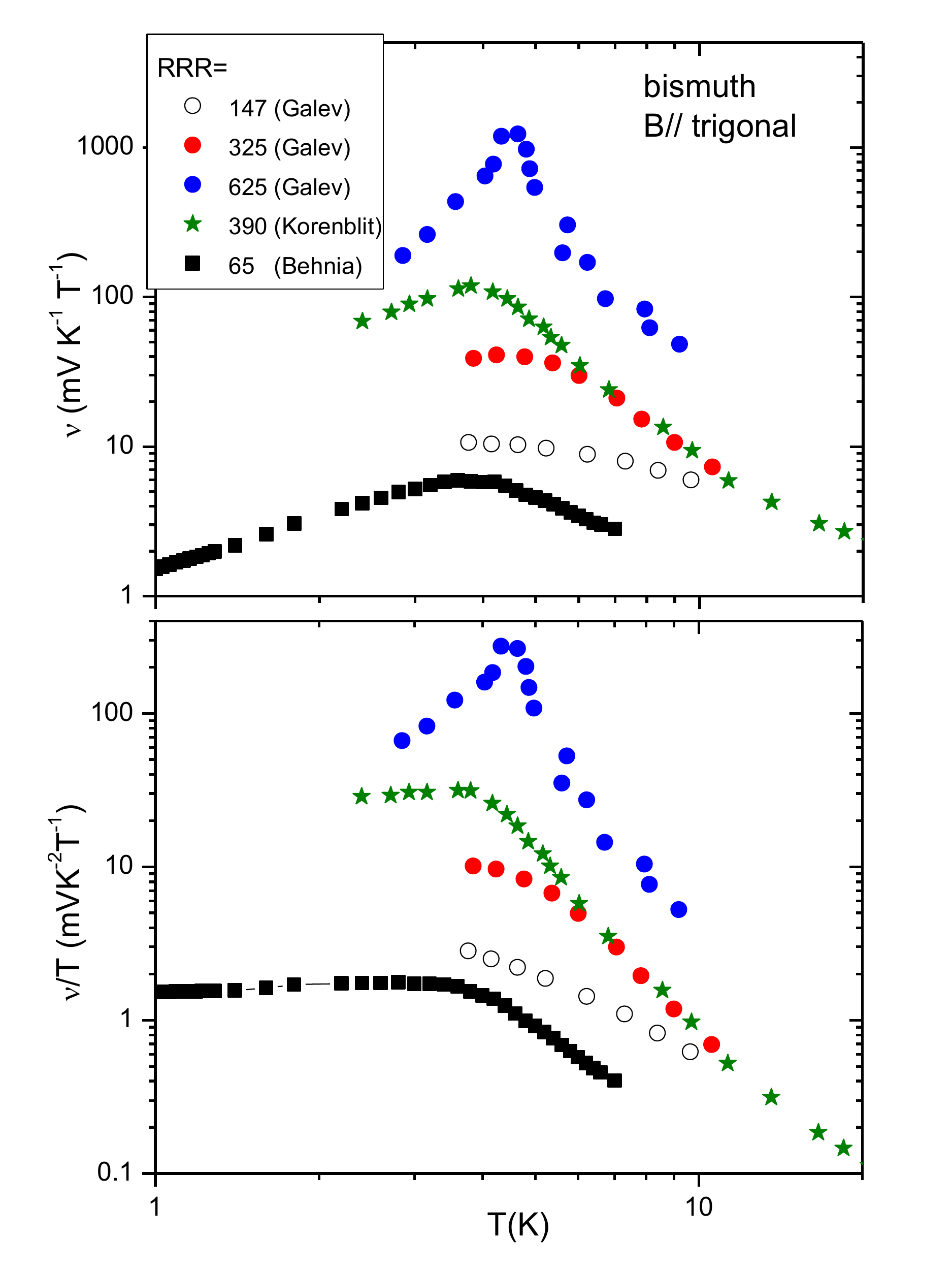}}
\caption{\label{Fig_Bi}Temperature dependence of the Nernst coefficient (Top) and the Nernst coefficient divided by temperature, $\nu/T$ (Bottom) in bismuth crystals reported by different authors\cite{korenblit1969,behnia2007,galev1981}. Note the difference in the Residual Resistivity Ratio (RRR= $\frac{\rho(300K)}{\rho(4.2K)}$) of the samples. The peak in the vicinity of 4K is believed to be caused by phonon drag. The finite $\nu/T$ resolved in the zero-temperature limit points to a purely diffusive component, which is larger in samples with larger RRR and therefore higher electron mobility.}
\end{center}
\end{figure}

In 2007, quantum oscillations were observed for the first time in an underdoped high-T$_{c}$ cuprate in presence of a strong magnetic field\cite{doiron2007}. Soon, it became clear that the quantized orbit is associated with a small Fermi surface resulting from reconstruction by an electronic order competing with superconductivity. The magnitude of the Nernst coefficient resolved at low temperature and high magnetic field was found to be in rather good agreement with the expected value, using the measured mobility of the carriers residing in this pocket and their Fermi energy \cite{chang2010,laliberte2011,chang2011}. The competing order has been identified as a charge order by a number of distinct experimental probes\cite{wu2011,ghiringelli2012,chang2012b}. The multiplicity of these electron pockets, their position in the Brillouin zone and the possible presence of hole pockets are still a subject of debate.

Fig. \ref{Fig_Nernst_EF} presents an update of a figure originally published in a previous review article\cite{behnia2009}. As seen in the figure, the new data confirms the already visible trend\footnote{ A figure which includes other cuprates and oxides can be found in ref.\cite{laliberte_PhD}}. Knowing the order of magnitude of the Fermi energy and the electronic mobility in a given system allows one to formulate an estimation for the magnitude of the low-temperature Nernst signal.

\begin{figure}
\begin{center}
{\includegraphics[width=10cm,keepaspectratio]{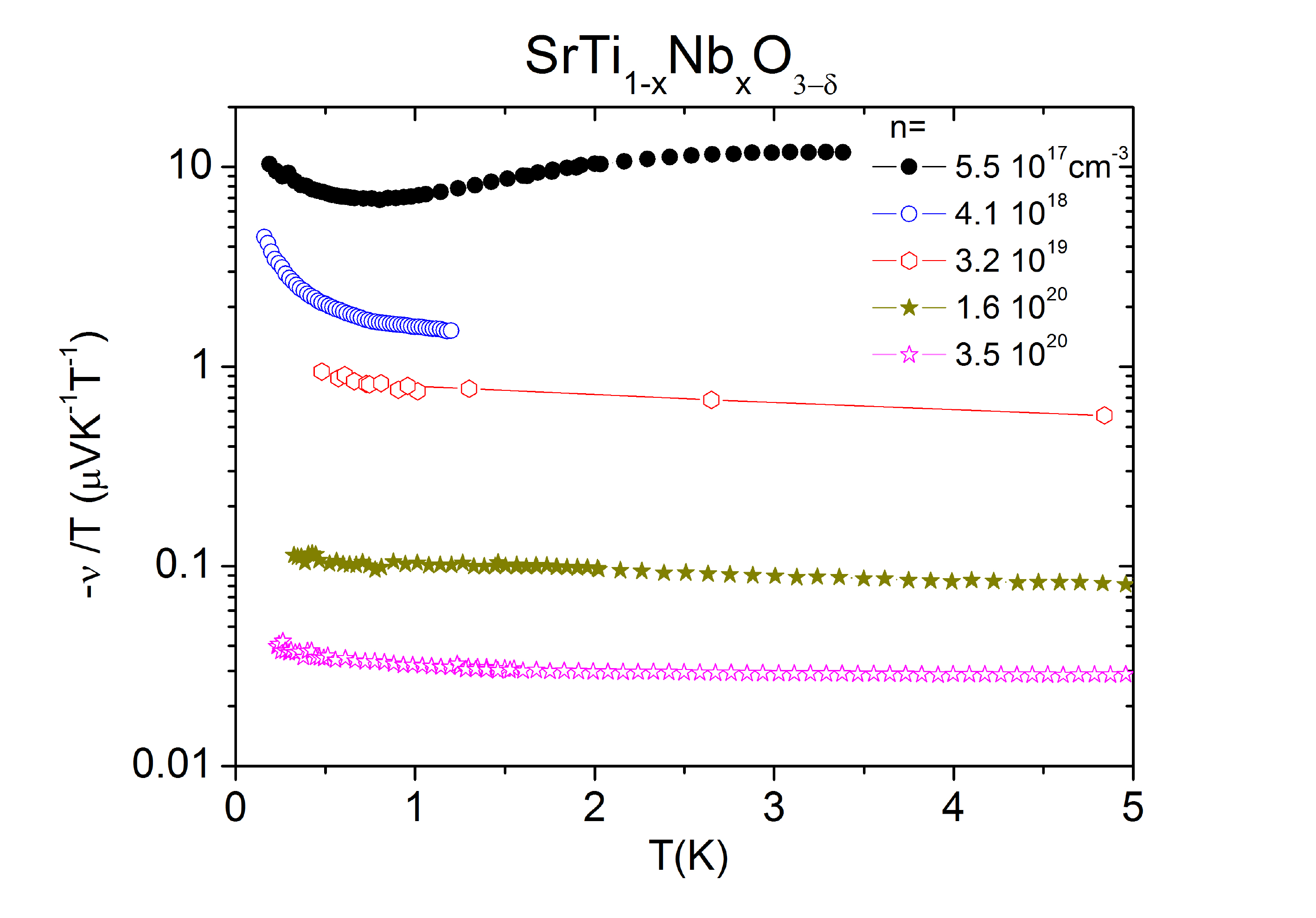}}
\caption{\label{nu_STO}Temperature dependence of the Nernst coefficient divided by temperature in crystals of n-doped SrTiO$_{3}$ with different carrier densities. As the carrier concentration decreases,  the Fermi energy lowers and the electron mobility increases. Both are expected to lead to a larger Nernst response in agreement with what is experimentally observed.}
\end{center}
\end{figure}

In the picture sketched above, one would expect to see that increasing quasi-particle mobility in a given system leads to an enhancement of its Nernst coefficient.  This has not been the subject of a systematic study. However, an examination of several papers on elemental bismuth is instructive. Back in 1969, Korenblit and co-workers\cite{korenblit1969} measured the Nernst coefficient of a bismuth single crystal  down to 2 K. The  Residual Resistivity Ratios (RRR= $\frac{\rho(300K)}{\rho(4.2K)}$) of the crystal was 390. They found that the Nernst coefficient peaks around 4K. This is the temperature at which the thermal conductivity (almost exclusively due to phonons) also peaks. Phonon drag  was plausibly invoked as the main source of the Nernst signal. A more recent study, published in 2007\cite{behnia2007} on a smaller single crystal  with a lower RRR found a still large but substantially lower Nernst signal. The extension of the data to lower temperatures allows one to clearly extract a finite T-linear component in the Nernst coefficient. This component cannot come from phonon drag (expected to follow a T$^{3}$ decay, much faster than T-linear). To these two reports, one can add a third mostly forgotten paper, published in 1981 by  Galev \emph{et al.}\cite{galev1981} measuring three different single crystals with different RRRs including one as high as 625. As illustrated in Fig.\ref{Fig_Bi}, putting all the data together, a correlation between high mobility and large  Nernst coefficient is clear. The phonon drag peak becomes remarkably sharp in the cleanest sample, as already underlined by Galev \emph{et al.}\cite{galev1981}. However, as seen in the bottom panel of the  figure, there is also a rigid upward shift in $\nu/T$ with increasing mean-free-path. This implies a positive correlation between the diffusive component of the Nernst signal and carrier mobility\footnote{The phonon-drag component of the Nernst response, which peaks at 4K is an interesting subject by itself and may point to an unusual resonant coupling between phonons and collective electronic excitations in  bismuth\cite{chudzinsky2015}.}.

In the case of degenerate semiconductors (sufficiently doped to be metallic), one can modify the Fermi energy of the mobile electrons by changing the dopant concentration. One expects that this would modify the longitudinal and transverse components of the thermoelectric response. In  the case of n-doped metallic SrTiO$_{3}$, experimental reports have documented the evolution of both the Seebeck \cite{okuda2001,cain2013}and  Nernst\cite{lin2013} coefficients with carrier concentration. The data reported in Ref.\cite{lin2013} on the Nernst coefficient is shown in Fig\ref{nu_STO}. The magnitude of the Nernst coefficient gradually increases as the density of carriers is lowered. Since lowering the carrier density pulls down the Fermi energy and pushes up the mobility at the same time, the drastic evolution of the amplitude of the Nernst coefficient is no surprise.

To summarize, the order of magnitude of the Nernst coefficients in metals in the low temperature limit seems to be well-understood. One shall not forget that at finite temperature phonon drag can give a large contribution to the Nernst signal. There are two recent examples.  Li$_{0.9}$Mo$_{6}$O$_{17}$ is a quasi-one-dimensional metal in which a large Nernst signal persists down to 35 K\cite{cohn2012}. Phonon drag is the most likely source of the observed signal (which was not measured below 10 K). In semiconducting FeSb$_{2}$, a large Nernst peak is observed around 7 K but rapidly vanishes at lower temperatures\cite{sun2013}. This is close to temperature at which the Seebeck coefficient and the thermal conductivity both peak. A recent theoretical work has identified phonon drag as the origin of the large thermoelectric response in this system\cite{battiato2015}.

\section{Review of experiments II: Short-lived Cooper pairs}
Much of the motivation for the recent research on the Nernst effect and its origins came from the observation  of a sizable Nernst signal above $T_c$ in underdoped cuprates\cite{xu2000,wang2006}. These early experiments were interpreted in the context of the ongoing debate on the origin of the pseudogap in cuprates.

According to BCS theory, cooling a superconductor below its superconducting transition temperature leads simultaneously to the formation of Cooper pairs and their condensation into a macroscopically coherent quantum state. However, Cooper pairs may also exist without macroscopic phase coherence, mostly as a consequence of thermal or quantum fluctuations of the Superconducting Order Parameter (SOP)\cite{fisher1991,kosterlitz1973,blatter1994,emery1994}. The magnitude of these fluctuations and their predominance in the phase diagram depends on various material-dependent parameters such as the concentration of random impurities, i.e. quenched disorder, dimensionality or the superconducting correlation length\cite{blatter1994}.

\begin{figure}
	\begin{center}
		{\includegraphics[width=12cm,keepaspectratio]{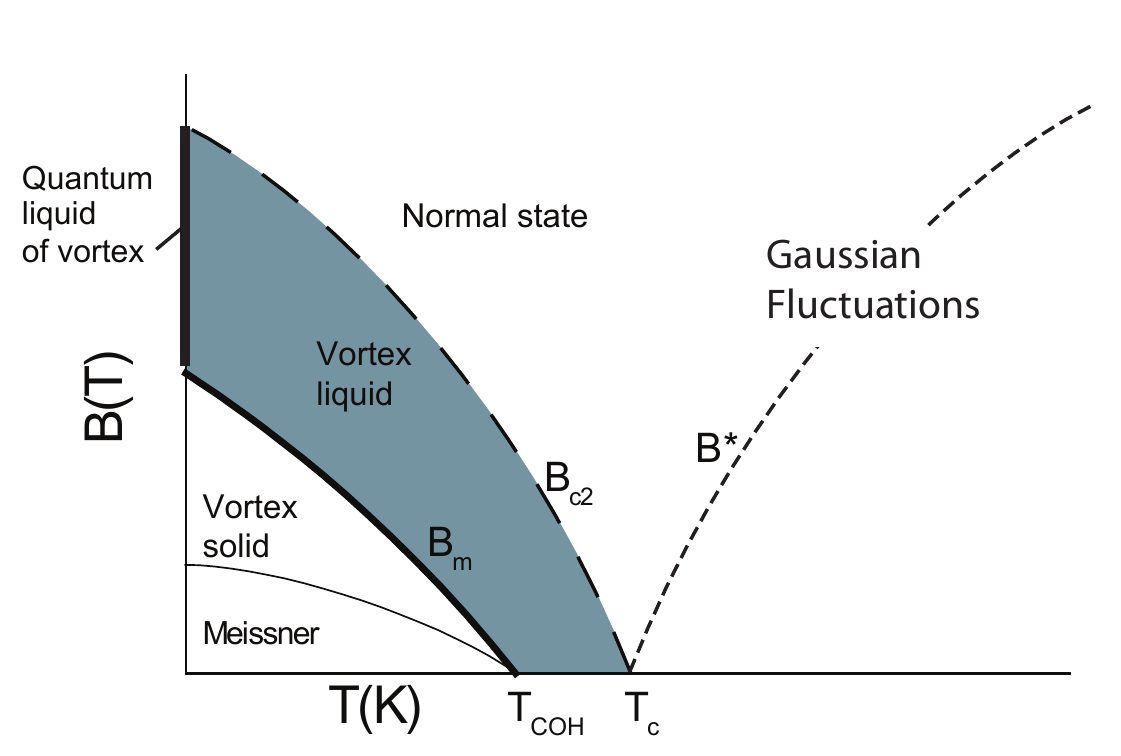}}
		\caption{\label{Fig:Diagramme_vortex} Theoretical phase diagram of a type II superconductor including the effects of thermodynamic and quantum fluctuations. Thermodynamic fluctuations of the phase of the order parameter can generate a fluid of vortex-anti vortex pairs even at zero magnetic field above the critical temperature. The quantum fluctuations can produce a quantum liquid of vortices at zero temperature in the vicinity of the upper critical field. The Nernst experiments could not distinguish between T$_{c}$ and T$_{coh}$.}
	\end{center}
\end{figure}

The vortex-liquid state, found in many conventional and non-conventional superconductors, is the most common regime of superconducting fluctuations displaying only phase fluctuations. This vortex fluid results from the melting of the vortex-solid above some characteristic magnetic field, $B_m$ \cite{fisher1991,blatter1994}, as a consequence of thermodynamic fluctuations of the phase of SOP.  As shown in the phase diagram sketched in Fig.~\ref{Fig:Diagramme_vortex}, this vortex fluid is separated from the normal state only by a crossover line representing the upper critical field $B_{c2}(T)$.

In high-T$_c$ superconductors, a combination of high temperature, small correlation length, large magnetic penetration depth and quasi-two-dimensionality, conspire to increase the effects of thermodynamic fluctuations and $B_m$ can be significantly smaller than the upper critical field $B_{c2}$. The observation of a pseudogap above $T_c$, in the underdoped region of their phase diagram, was interpreted by many resesearchers as the possible signature of a superconducting phase diagram with two characteristics temperatures. In this scenario, the higher characteristic temperature, where the pseudogap forms in the electronic spectrum, would correspond to formation of Cooper pairs and the lower characteristic temperature would correspond to the \emph{actual} transition toward the phase-coherent superconducting state\cite{lee2006}. Between these two temperatures exists a regime with only phase fluctuations of the SOP, which is fundamentally different from the regime of Cooper pair fluctuations as described in the context of Ginzburg-Landau theory\cite{larkin2009}. Let us recall, that in this last theory only one single critical temperature, $T_c$, is required to describe the fluctuations. Remarkably, in the Ginzburg-Landau theory, there is no upper temperature limit for the existence of these fluctuations; they are expected to survive far above $T_c$. In contrast, the regime of phase fluctuations exists only between two characteristics temperatures.

In the context of cuprates physics, Emery and Kivelson \cite{emery1994} extended the concept of phase-coherence temperature introduced by Berezinski, Kosterlitz and Thouless\cite{kosterlitz1973,ambegaokar1980} and suggested that, for any superconductor, vortex-antivortex pairs should appear spontaneously when the thermal energy, $k_BT$, is larger than the energy cost for their formation; this energy cost results from the kinetic energy associated with superfluid flow around the vortices. This defines a characteristic temperature for phase coherence, $T_{COH}$, above which spontaneous nucleation of vortices is possible. In conventional superconductors, this coherence temperature largely exceeds $T_{BCS}$, the Cooper pair forming
temperature, and superconducting fluctuations exist only as Cooper pairs fluctuations. In contrast, for low density superconductors, $T_{COH}$ may become smaller than $T_{BCS}$. This implies that the superconducting transition is controlled by the superfluid density.

%In the context of cuprates physics, this could provide an explanation of
%the Uemura plot \cite{Uemura1989} where $T_c$ is found to scale with
%the magnetic penetration depth which is inversely proportional to
%superfluid density.

%In addition to quenched disorder and thermal fluctuations,
%quantum fluctuations of the SOP provide another origin for the
%\emph{quantum} melting of the vortex solid. This leads to a phase
%diagram as shown in panel c) of figure \ref{fig:fig1}, where a quantum
%liquid of vortices is expected in the zero-temperature limit,
%separated from the superconducting state by a second order
%transition whose critical behavior is controlled by quantum
%fluctuations \cite{Sondhi1997}. Fine-tuning of the transition can be
%achieved either by applying a perpendicular magnetic
%field \cite{Hebard1990,Paalanen1992,Yazdani1995,Ephron1996,Markovic1998a,Markovic1998b,Gantmakher2000,
%	Bielejec2002,Sambandamurthy2004,Aubin2006} or by varying the sheet
%resistance $R_{\square}$ of the films -- using film
%thickness \cite{Jaeger1989,Markovic1999,Kikuchi2008} or electrostatic
%field \cite{Parendo2005}.

The most important experimental support for the existence of a regime of phase fluctuations in cuprates came from the  Nernst studies on underdoped cuprates such as La$_{2-x}$Sr$_xCuO_4$(LSCO)\cite{xu2000,wang2006}. The interpretation of these experiments was based on two assumptions: first, that the fermionic quasiparticles have a negligible contribution to the Nernst signal and second, that Gaussian superconducting fluctuations cannot produce a Nernst signal well above the critical temperature  ($T>2\times T_c$). Both these assumptions were proved to be wrong.

As discussed in the previous section, the quasi-particle contribution to the Nernst response can be  very large. In the case of cuprates, the quasi-particle contribution to the Nernst coefficient is far from negligible\cite{cyrChoiniere2009,tafti2014}. Furthermore, as described below, in a conventional superconductor, namely Nb$_x$Si$_{1-x}$, fluctuating Cooper pairs can produce a Nernst signal up to a high temperatures ($T\simeq 30\times T_c$) and high magnetic field ($H\simeq 4\times H_{c2}$)\cite{pourret2006,pourret2007}. Close to $T_c$, the data on Nb$_xS$i$_{1-x}$ and also  on the amorphous superconductor InO$_x$\cite{spathis2008,pourret2009}, was found to be in quantitative agreement with the theory of the Nernst signal generated by Gaussian fluctuations conceived by Ussishkin, Sondhi, Huse(USH)\cite{ussishkin2002}. The validity of this theory was restricted to low magnetic field and the vicinity of critical temperature. Following the experiments on Nb$_xS$i$_{1-x}$, two groups developed theoretical calculations valid at higher temperature and higher magnetic field. The results \cite{serbyn2008,michaeli2009,michaeli2009a} were found to be in quantitative agreement with experimental data.

Following these developments, Taillefer and collaborators revisited the Nernst effect in cuprates and clearly identified the Nernst signal due to Cooper pair fluctuations in hole-doped  La$_{1.8-x}$Eu$_{0.2}$Sr$_xCuO_4$ (Eu-LSCO)\cite{chang2012} and electron-doped cuprate Pr$_{2-x}$Ce$_x$CuO$_4$ (PCCO)\cite{tafti2014}. They concluded that no contribution other than the quasiparticles and fluctuating Cooper pairs could be identified either in the underdoped or in the overdoped regimes. Consequently, it is fair to say that experimental support for phase-only superconducting fluctuations in cuprates has faded away. We review below the characteristic signatures of the Nernst signal generated by Cooper pair fluctuations, as observed in Nb$_x$Si$_{1-x}$\cite{pourret2006,pourret2007} and cuprates\cite{chang2012,tafti2014}.

As discussed in section 2, $\alpha_{xy}$ of fermionic quasiparticles scales with their mean free path, but in the case of short-lived Cooper pairs it only depends on the superconducting correlation length. In an amorphous superconducting film where the elastic mean free path is of the order of the inter-atomic distance, the quasiparticle contribution becomes extremely small while the contribution of the fluctuating Cooper pairs can be large if the superconductor has a long correlation length.

The Nernst signal due to short-lived Cooper pairs was first identified in amorphous thin films of Nb$_{0.15}$Si$_{0.85}$\cite{pourret2009,pourret2006,pourret2007} and InO$_x$\cite{spathis2008}. In those films as well as in cuprates, the transverse Hall conductivity $\sigma_{xy}$ is small. As mentioned above, this simplifies the relationship between the Nernst coefficient, $\nu$ and  $\alpha_{xy}$:

\begin{equation}
\nu\approx\frac{\alpha_{xy}}{B\sigma_{xx}}
\end{equation}

\begin{figure}[ht!]
	\begin{center}
		\includegraphics[width=13cm]{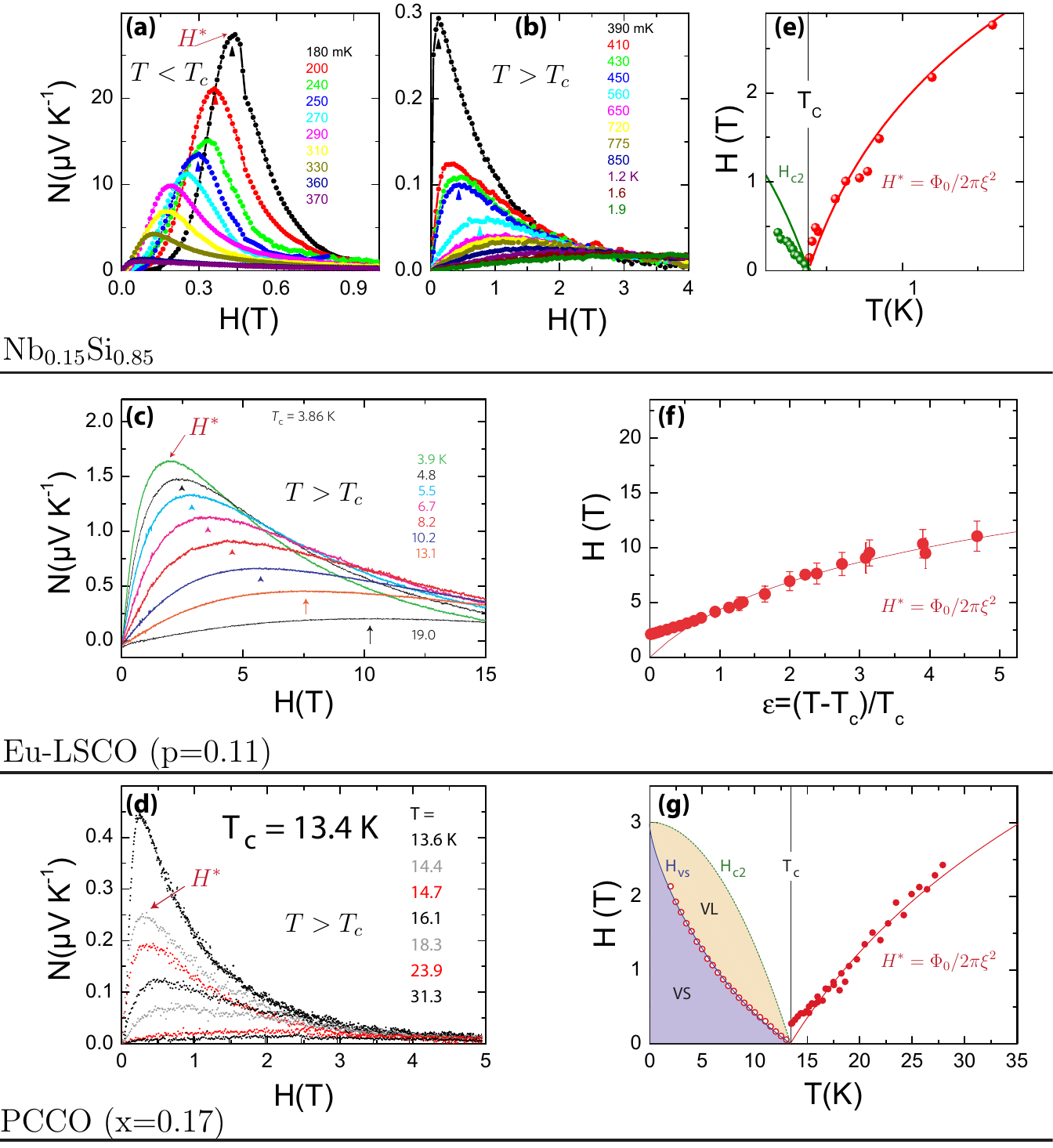}
		\caption{\label{Fig:NernstField}  Nernst signal in Nb$_{0.15}$Si$_{0.85}$ measured below (a) and and above (b) $T_c$. (c) Nernst signal in Eu-LSCO measured above $T_c$. (d) Nernst signal in PCCO measured above $T_c$. Maxima at $B^*$ are indicated by arrows. Below $T_c$, $B^*$ increases with decreasing temperature. Above $T_c$, the temperature dependence of $B^*$ is reverted, it increases with increasing temperature. e) The field scale $B^*$ as a function  of temperature for Nb$_{0.15}$Si$_{0.85}$. $B^*$ vanishes at $T_c$. Below $T_c$, this is the field at which the vortex  Nernst signal peaks. Above $T_c$, it represents the ``ghost critical field'' (GCF).   f) $B^*$ as function of reduced temperature in Eu-LSCO ~\cite{chang2012}. g) The field scale $B^*$ as a function of temperature in PCCO.~\cite{tafti2014}. In all three systems, the GCF follows  $B^*=\frac{\Phi_0}{2\pi \xi^2}$.
			}
	\end{center}
\end{figure}

Fig.\ref{Fig:NernstField} shows the magnetic field dependence of the Nernst signal for Nb$_{0.15}$Si$_{0.85}$\cite{pourret2009,pourret2006,pourret2007}, Eu-LSCO\cite{chang2012} and PCCO\cite{tafti2014}. In the normal state , the field dependence of the Nernst signal displays a maximum at  $B^*$. The magnitude of $B^*$ increases with increasing temperature.

In the case of Nb$_{0.15}$Si$_{0.85}$, the vortex-induced Nernst signal measured below $T_c$ is also shown in the same figure. It also shows a maximum as a function of magnetic field. In contrast to the normal state, the position of the maximum $B^*$ shifts toward higher magnetic fields upon decreasing the temperature. This is not surprising, since in the superconducting state, all characteristic fields associated with superconductivity, the upper critical field  $B_{c2}$ and the vortex melting field $B_m$, are expected to increase with decreasing temperature. Plotting the position of $B^*$, above and below $T_c$, on the phase diagram of Fig.\ref{Fig:NernstField}, one can see that $B^*$ vanishes at $T_c$. This observation clearly indicates that the nature of superconducting fluctuations at the origin of the Nernst signal observed above $T_c$ is fundamentally distinct than below $T_c$. Below $T_c$, the Nernst signal is generated by long-lived vortices of the vortex liquid. On the other hand, above $T_c$, it is generated by fluctuating Cooper pairs.

%\begin{figure}[ht!]
%	\begin{center}
%	\includegraphics[width=12cm]{ContributionNernst}
%	\caption{\label{Fig:ContributionNernst} Temperature dependence of the Nernst coefficient measured at two distinct values of the magnetic field, $B=2~T$ and $B=0.04~T$. The measured Nernst coefficient is much higher than the quasiparticle contribution $S \tan{\theta}\times 2000$.}
%	\end{center}
%\end{figure}
%
%An estimation of the quasiparticule contribution can be obtained from a measure of the Seebeck coefficient and the Hall angle. A comparison with the Cooper pair contribution, \ref{Fig:ContributionNernst}, shows that the quasiparticles provide a negligeable contribution to the Nernst signal up to very high temperature above $T_c$.

Fluctuating Cooper pairs correspond to spatial and temporal fluctuations of the SOP, $\Psi(x,t)$, and are described by the Ginzburg-Landau theory\cite{larkin2009}. The typical size of these superconducting fluctuations is set by the correlation length, $\xi$. It sets the characteristic length on which the correlation function $<\psi(x_0)\psi(x_0-x)>$ decreases to zero. Upon cooling, this correlation length increases and diverges at the approach of the superconducting transition following $\xi=\xi_0 \varepsilon^{-1/2}$ where $\varepsilon=\ln(T/T_c)$. At the microscopic
level, these fluctuations correspond to short-lived Cooper pairs whose life-time is controlled by their decay into free electrons :

\begin{equation}
\tau=\frac{\pi\hbar}{8k_BT_c}\varepsilon^{-1}
\end{equation}

These fluctuations give rise to the phenomena of paraconductivity \cite{glover1967} and fluctuating diamagnetism \cite{gollub1973}. Normal quasi-particles contribute significantly to conductivity and magnetic susceptibility. Therefore, the sensitivity of these probes to superconducting fluctuations is limited to a narrow region close to the superconducting transition \cite{skocpol1975}. This is to be contrasted with the Nernst experiments. In amorphous superconducting films, the short elastic mean free path of a few Angstroms, drastically weakens the contribution of free electrons to the Nernst response, which becomes orders of magnitude lower than the signal due to superconducting fluctuations. This is why the Nernst signal generated by short-lived Cooper pairs can be detected up to very high temperatures ($30\times T_c$) and high magnetic field ($4\times B_{c2}$), deep inside the normal state\cite{pourret2006,pourret2007}. Furthermore, because of this weak contribution of normal quasiparticles excitations, a direct and unambiguous comparison of the data with theory becomes possible.

\begin{figure}[ht!]
	\begin{center}
		\includegraphics[width=14cm]{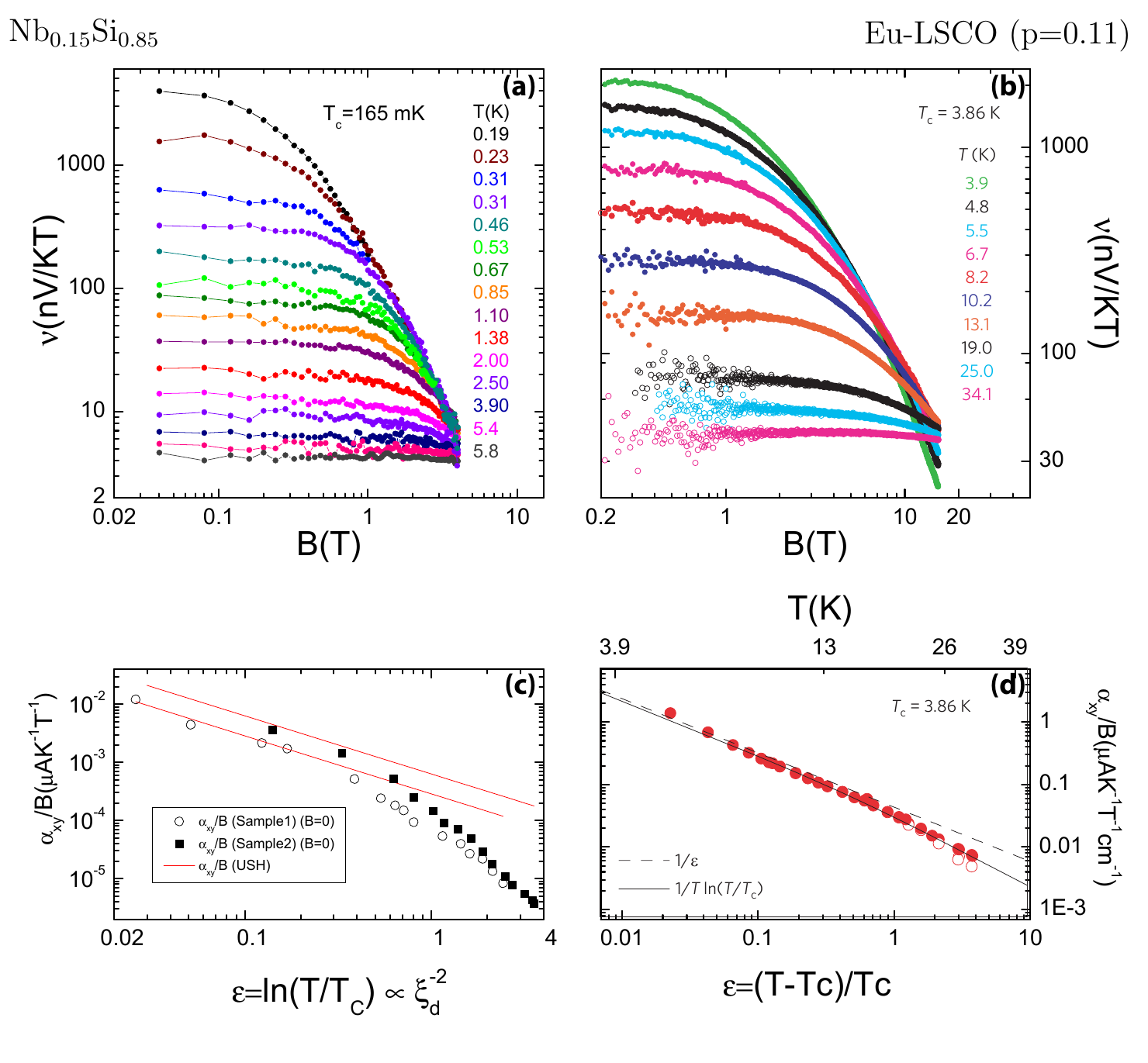}
		\caption{\label{Fig:Nernst_Coefficient} Nernst coefficient, $\nu$ ,as function of magnetic field in Nb$_{0.15}$Si$_{0.85}$ (a) and Eu-LSCO (b), measured at temperatures exceeding  $T_c$.  The two systems display a similar behavior. At low magnetic field  ($B<B^*$),  the Nernst coefficient is independent of magnetic field with a magnitude set by the temperature-dependent correlation length. At high magnetic field ($B>B^*$), the Nernst coefficient becomes independent of temperature with a magnitude determined by the magnetic length.
			$\frac{\alpha_{xy}}{B}=\sigma_{xx}\nu$ in the zero-field limit extracted from the measured Nernst coefficient and conductivity in  Nb$_{0.15}$Si$_{0.85}$ (c) and Eu-LSCO (d).
			The magnitude of this coefficient close to $T_c$ can be perfectly described by Eq.~\ref{eq:USH} with the correlation length as the only variable parameter.}
	\end{center}
\end{figure}

Treating the fluctuations of the SOP in the Gaussian approximation, Ussishkin \emph{et al.}\cite{ussishkin2002} obtained a simple analytical formula, valid close to $T_c$ and restricted to the zero-magnetic field limit:

\begin{equation}\label{eq:USH}
\frac{\alpha^{SC}_{xy}}{B}= \frac{1}{6\pi}\frac{k_{B} e^2}{\hbar^2}\xi^{2}
\end{equation}

\begin{figure}[ht!]
	\begin{center}
		\includegraphics[width=14cm]{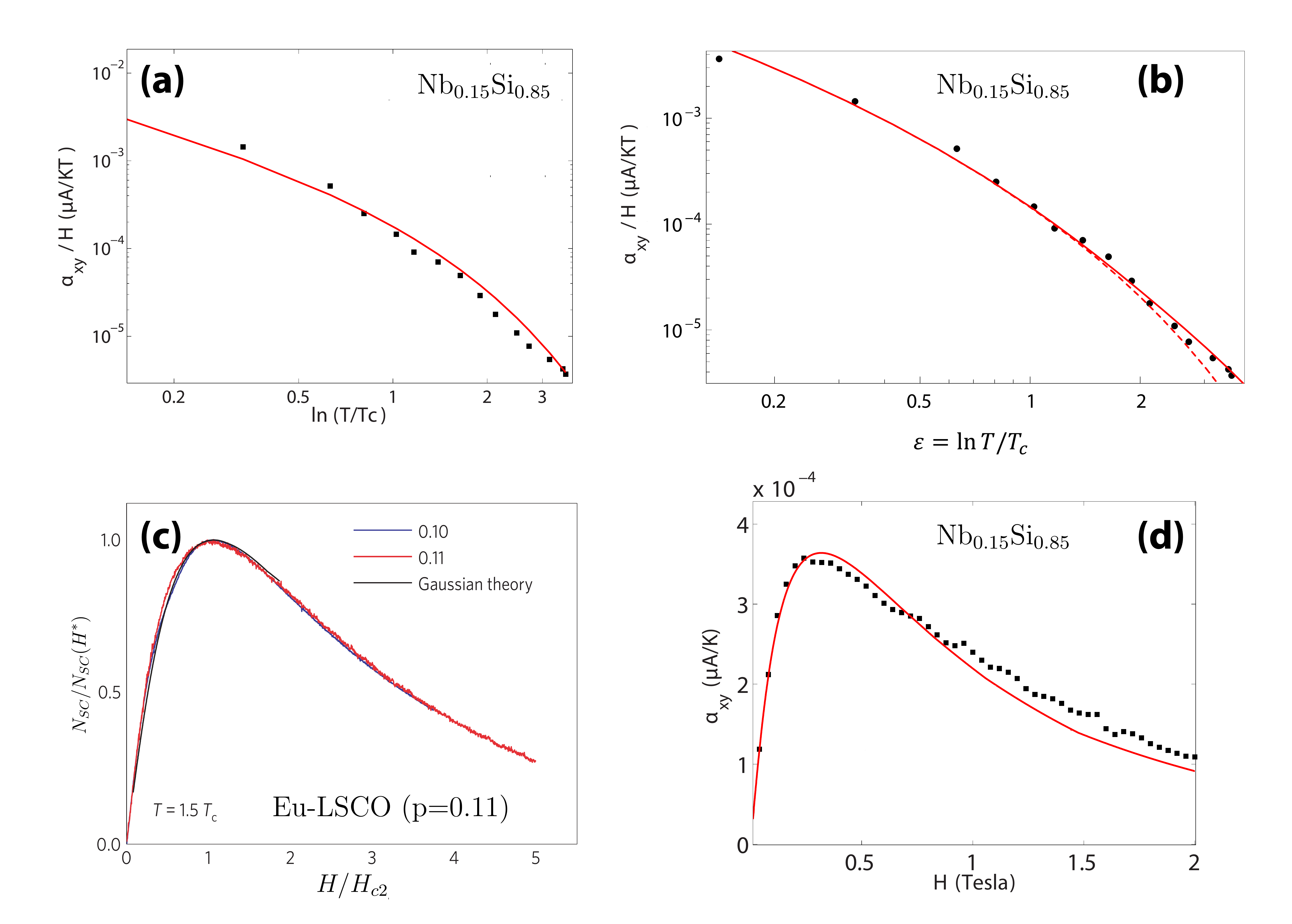}
		\caption{\label{Fig:TestTheorie} Temperature dependence of  $\frac{\alpha_{xy}^{SC}}{B}$ for Nb$_{0.15}$Si$_{0.85}$ compared with a theory by Serbyn {\it et al.}\cite{serbyn2008} (a) and with the theory by Michaeli {\it et al.}\cite{michaeli2009,michaeli2009a} ( b). These theories describe the experimental data well above T$_{c}$. Magnetic field dependence of $\frac{\alpha_{xy}^{SC}}{B}$ for Nb$_{0.15}$Si$_{0.85}$ (c) and Eu-LSCO (d) compared with  theory by Michaeli {\it et al.}\cite{michaeli2009,michaeli2009a}. This theory successfully describes the experimental data even at high magnetic field and reproduces the maximum observed at $B^*$.}
	\end{center}
\end{figure}

According to Eq.~\ref{eq:USH}, the coefficient $\alpha^{SC}_{xy}/B$ is independent of magnetic field. Figure~\ref{Fig:Nernst_Coefficient} shows its measured value for Nb$_{0.15}$Si$_{0.85}$ and Eu-LSCO, where it can be indeed seen that this is independent of magnetic field at low magnetic field.

From those plots, the value of $\frac{\alpha^{SC}_{xy}}{B}$ in the zero magnetic field limit $(B\rightarrow 0)$ is extracted and compared to what is expected according to Eq.\ref{eq:USH}. As shown in Fig.~\ref{Fig:Nernst_Coefficient}, a quantitative agreement with the theoretical prediction is found close to $T_c$. At high temperature, the data deviate from the USH theoretical expression, which is valid only close to $T_c$.  Further theoretical works have extended the calculations to higher temperature and magnetic field\cite{serbyn2008,michaeli2009,michaeli2009a} and have been found to be in quantitative agreement with those data as well, as shown in Fig.~\ref{Fig:TestTheorie}.

From this analysis in the zero magnetic field limit, one understands that the amplitude of the Nernst coefficient is set by a single characteristic length, the size of superconducting fluctuations\cite{pourret2007,spathis2008}. In the zero-field limit, this size is set by the correlation length, $\xi$. In the high field limit, the size of superconducting fluctuations is set by the magnetic length, (here defined as $\ell_B=(\hbar/2eB)^{1/2}$; note that $e$ of quasi-particles is replaced with $2e$ of Cooper pairs). This happens when this length becomes shorter than the zero-field correlation length.

The field-induced shrinking of superconducting fluctuations is well-known thanks to studies of fluctuating diamagnetism in both low-temperature superconductors\cite{gollub1973} and high-T$_{c}$ cuprates \cite{carballeira2000}. In the low field limit, the magnetic susceptibility should be independent of the magnetic field. This is the so-called Schmidt limit\cite{schmid1969}. Experimentally, the magnetic susceptibility is  observed to decrease with  magnetic field, following the Prange's formula\cite{prange1970},  an exact result within the Ginzburg-Landau formalism. At high magnetic field, the superconducting fluctuations are described as evanescent Cooper pairs arising from free electrons with quantized cyclotron orbits\cite{skocpol1975}.

As a consequence of this phenomenon, at a given temperature  above T$_c$, when the magnetic field exceeds $B^*=\phi_0/2\pi{\xi}^2$, the typical size of superconducting fluctuations decreases from its zero-field value (set by  $\xi(T)=\xi_0 \varepsilon^{-1/2}$) towards the magnetic length value $\ell_B$.  This characteristic field was identified first by Kapitulnik, Palevski and Deutscher  studying the magnetoresistance of mixture films of InGe\cite{kapitulnik1985}. It mirrors, above $T_c$, the upper critical field below $T_c$. Therefore, these authors dubbed it the ``Ghost Critical Field''.

%\begin{figure}[ht!]
%	\begin{center}
%		\includegraphics[width=14cm]{Nernst_Diagram}
%		\caption{\label{Nernst_Diagram} Cartographie couleur du signal Nernst, échelle logarithmique, en fonction du champ magnétique et de la température. Les cercles noirs pleins indiquent le champ critique fantôme $B^*$, les carrés vides indiquent le champ critique $B_{c2}$. Constatez l'évolution symétrique de ces deux échelles de champ magnétique de par et d'autre de $T_c$. La ligne continue représente le champ magnétique déterminée par la longueur de corrélation de Ginzburg-Landau, $B^*=\phi_0/2\pi\xi^2$. Cette figure est extraite de Ref.~\cite{Pourret2007}.}
%	\end{center}
%\end{figure}

\begin{figure}[ht!]
	\begin{center}
		\includegraphics[width=14cm]{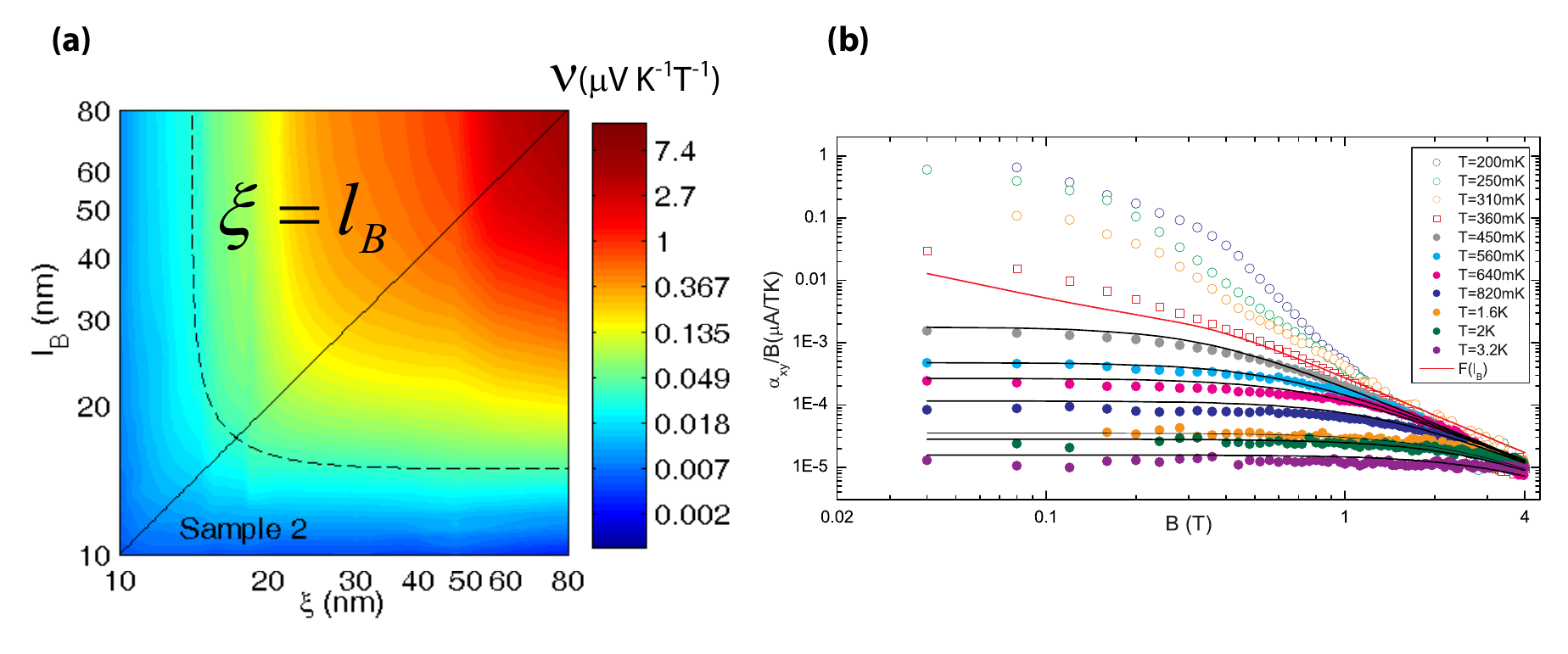}
		\caption{\label{Fig:OneLengthGraph} a) Color map (in logarithmic scale)  of the Nernst coefficient  as a function of the magnetic length $\ell_B$ and correlation length $\xi$ for Nb$_{0.15}$Si$_{0.85}$. One can see that the Nernst coefficient is symmetric with respect to the diagonal, $\ell_B=\xi$. b) Log-Log plot of $\alpha_{xy}^{SC}$ as function of magnetic field for different temperatures in the range 0.2~K$<T<$3.2~K. The function $F(\xi)=[\alpha_{xy}/B)](\xi)$ is extracted from the temperature dependence of the Nernst coefficient in the zero-field limit. Plotting this curve as function of magnetic field, using  $\ell_B$ as the argument, gives the red curve shown. One can see that this curve describes properly the Nernst coefficient as function of magnetic field measured at the critical temperature. Indeed,  since the correlation length $\xi$ becomes infinite at the critical temperature, the size of superconducting fluctuations is set by the magnetic length at any magnetic field.}
	\end{center}
\end{figure}

Above $T_c$, this crossover is responsible for the observed maximum in the field dependence of the Nernst signal, shown Fig.~\ref{Fig:NernstField}. Upon increasing the magnetic field, the Nernst signal increases linearly with field, reaches a maximum at $B^*$ and decreases afterwards. Our previous papers\cite{pourret2007,spathis2008} presented detailed arguments demonstrating that the Nernst coefficient is set by the size of superconducting fluctuations and that $B^*$, the field at which the Nernst signal peaks in the normal state is close the ghost critical field ($GCF =\frac{\phi_{0}}{2\pi \xi_{d}^2}$) \footnote{Recently, Kavokin and Varlamov\cite{kavokin2015} argued that far above T$_{c}$,  the temperature dependence of B$^*$ is expected to be more complicated than what has been assumed. It remains an experimental fact that the field at which the Nernst effect peaks above T$_{c}$ mirrors within a good precision the evolution of H$_{c2}$ below T$_{c}$.}. Let us now summarize the main points.

\begin{itemize}
\item At low magnetic field, the magnitude of the Nernst coefficient depends only on the temperature and is independent of the magnetic field. This is because when $\ell_B>\xi$, the size of the superconducting fluctuations is set by the temperature-dependent correlation length $\xi$ (See Fig.~\ref{Fig:Nernst_Coefficient}).

\item Above $T_c$, both the magnitude and the temperature dependence of $B^*$ are set by the Ginzburg-Landau correlation length, $\xi_{d}=\frac{\xi_{0d}}{\sqrt{\varepsilon}}$ through the relation $B^{*}=\frac{\phi_{0}}{2\pi \xi_{d}^2}$ where $\phi_{0}$ is the flux quantum and $\varepsilon=\ln{\frac{T}{T_{c}}}$ (See Fig.~\ref{Fig:NernstField}). For Eu-LSCO and PCCO, B$^{*}$ deviates from this relation when $\varepsilon<.5$ and remains finite as $\varepsilon\rightarrow 0$. This has been attributed to the divergence of the paraconductivity in the limit  $\varepsilon\rightarrow 0$\cite{chang2012,tafti2014}. In this limit, the Nernst signal $N(\varepsilon)=\alpha_{xy}(\varepsilon)/\sigma(\varepsilon)$ is the ratio of two diverging quantities. Therefore,  a saturation of $N(\varepsilon)$ at low $\varepsilon$ is not surprising.

\item At high magnetic field, $B>B^*(T)$, the  Nernst data collapse on to a weakly temperature-dependent curve. Indeed, when $\ell_B<\xi$, the size of superconducting fluctuations is set by the magnetic length, which is obviously independent of temperature (See Fig.~\ref{Fig:Nernst_Coefficient}).

\item As shown in Fig.\ref{Fig:OneLengthGraph}, when one substitutes temperature and magnetic field by their associated length scales, the zero-field superconducting correlation length, $\xi(T)$ and the magnetic length, $\ell_B(B)$,  the Nernst coefficient for Nb$_{0.15}$Si$_{0.85}$, becomes symmetric with respect to the diagonal $\xi(T)=\ell_B$. This shows that the
	Nernst coefficient depends only on the size of superconducting fluctuations, no matter what sets it, the magnetic length or the correlation length.

\item The knowledge of the temperature dependence of the Nernst coefficient in the limit of zero magnetic field  is sufficient to describe the evolution of the Nernst coefficient with the amplitude of the magnetic field at $T=T_c$. Indeed, at T=T$_c$, the correlation length diverges and therefore for any amplitude of the magnetic field $\ell_B<(\xi\rightarrow \infty)$. Knowing the evolution of the Nernst coefficient as function of the correlation length (obtained from its temperature dependence in the zero-field limit), we can deduce its field dependence using the magnetic length (See Fig.~\ref{Fig:OneLengthGraph}).
\end{itemize}

\section{Review of experiments III: Mobile vortices }
\begin{figure}
\begin{center}
\resizebox{!}{0.7\textwidth}{\includegraphics{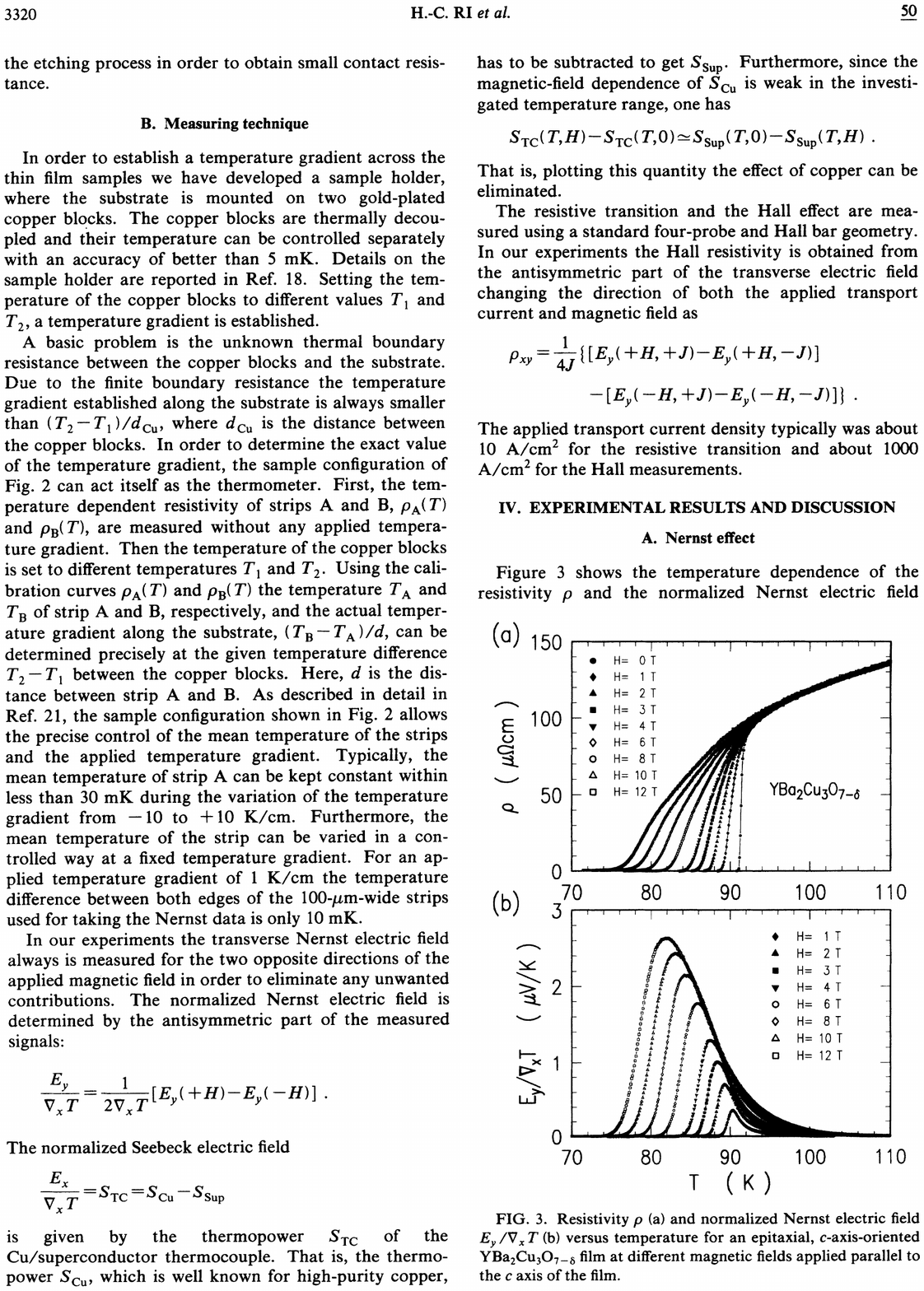}}
\caption{\label{Fig_Nernst_vortex_YBCO} Resistivity (top) and Nernst coefficient(bottom) in optimally-doped YBa$_{2}$Cu$_{3}$O$_{7-\delta}$ in the vicinity of the superconducting transition. The application of magnetic field broadens the  resistive transition as a consequence of vortex flow caused by the Lorenz force caused by the charge current. Magnetic field also generates a finite Nernst signal. At each magnetic field, the Nernst signal peaks at the middle of the resistive transition. It vanishes both at low temperature when the vortices can no more move and above T$_{c}$ where there are no more vortices left. Note, however, the presence of a fluctuating tail above T$_{c}$ (After Ref.\cite{ri1994}). }
\end{center}
\end{figure}
Below its critical temperature, a superconductor does not let the magnetic field penetrate except for a thin layer called the penetration depth. This Meissner state is eventually destroyed by a sufficiently strong magnetic field. In most superconductors (called type II for historical reasons), there is an intermediate phase between the Meissner state and the normal phase. This is the so-called mixed phase, emerging above a critical threshold, known as the first critical field (H$_{c1}$). The magnetic field then enters the system as a set of tiny filaments called  superconducting vortices. They have a normal core, which extends over the superconducting coherence length, and a periphery of whirling Cooper pairs, extending over the penetration depth, two material-dependent length scales. Because of the latter, each vortex carries a quantum of magnetic flux. As a consequence of the former, it is a reservoir of entropy. The combination of these two properties make vortices a potential source of a Nernst signal.

One can apply a force on vortices either by injecting an electric current or by imposing a thermal gradient. In the first case, the force is due to the Lorenz force between a moving electric charge and a magnetic field. In the second case, it is  because the normal core of a vortex has more entropy than the superconducting background, and the second law of thermodynamics will push it from hot to cold. Vortex movement generates an electric field as one would expect for a moving magnetic flux. In contrast to the electric field generated by the flow of normal quasi-particles, the electric field caused by the vortex movement is caused by the condensate itself and not an external perturbation and is not cancelled by Cooper pairs.

Thus, when superconducting vortices are mobile, one expects that a longitudinal thermal gradient generates a transverse electric field, in other words, a finite Nernst signal. Bridgman relation between Nernst and Ettingshausen coefficients implies that there should also be a finite Ettingshausen effect. This happens because injecting charge current can lead to vortex movement as a consequence of Lorenz force. Such mobile vortices carry their entropy  and generate a thermal gradient along their trajectory perpendicular to the current density vector. This  produces an Ettingshausen signal. Early experiments, performed on magneto-thermoelectricity of type II superconductors in their mixed state, detected both Nernst\cite{lowell1967,vidal1973} and Ettingshausen\cite{solomon1967} effects (See ref. \cite{huebener1972} for a review).

The discovery of high-temperature superconductors in 1986 opened a new era in the study of vortex dynamics. In these superconductors vortices are mobile in an extended region of the (field, temperature) plane\cite{blatter1994}. Following a pioneer experiment detecting the Ettingshausen signal\cite{palstra1990}, numerous experiments on thermally-induced movement of vortices followed (See ref. \cite{huebener1995} for a review). A typical set of data on vortex Nernst signal in cuprates is shown in Fig.~\ref{Fig_Nernst_vortex_YBCO}.

\begin{figure}
\begin{center}
\resizebox{!}{0.7\textwidth}{\includegraphics{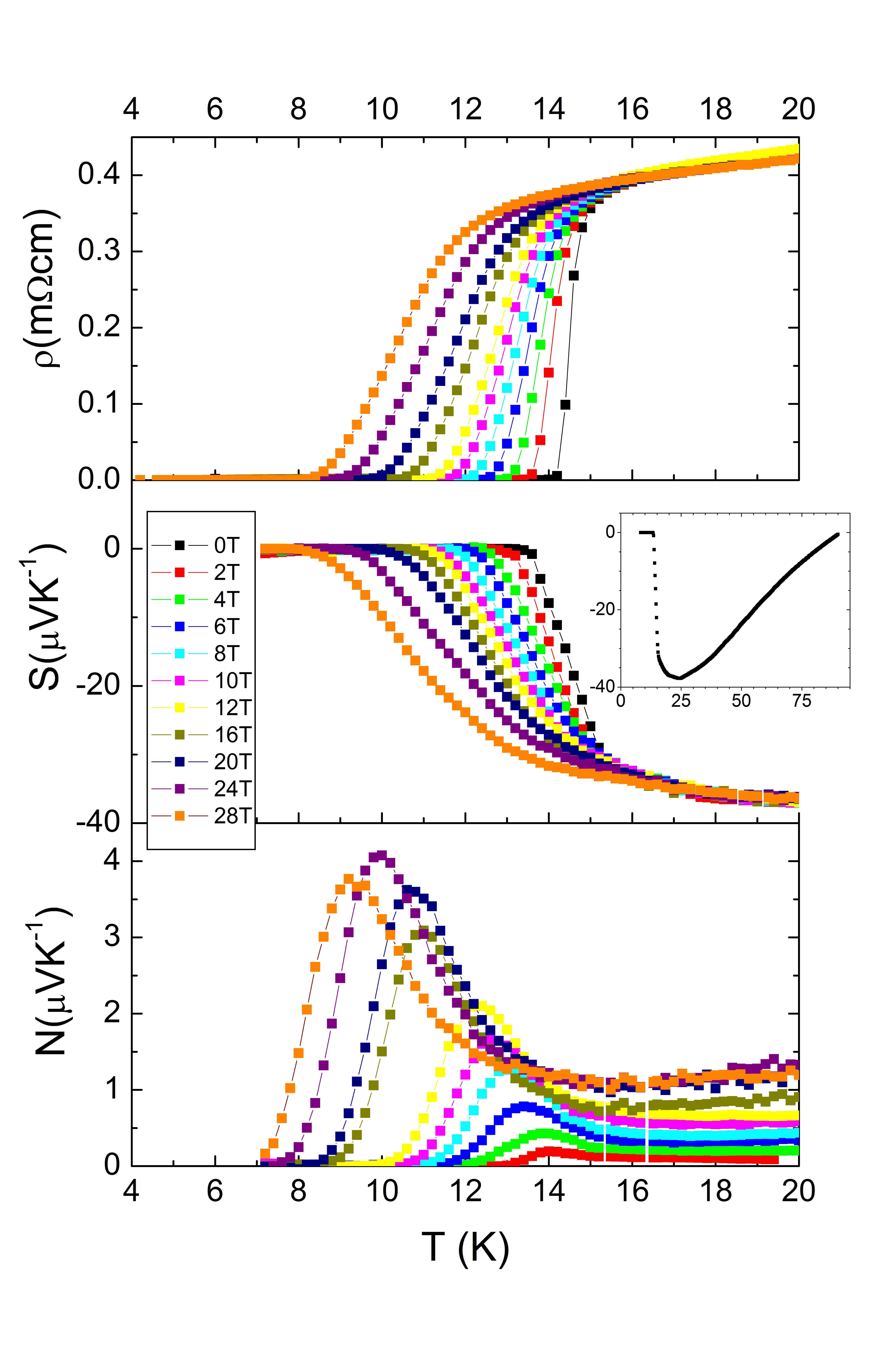}}
\caption{\label{Fig_Nernst_vortex_FeSeTe} Temperature dependence of the resistivity (top), the Seebeck coefficient (middle), and the Nernst coefficient (bottom) for different magnetic fields in
Fe$_{1+y}$Te$_{0.6}$Se$_{0.4}$. The inset shows the temperature dependence of the zero-field Seebeck coefficient up to 90 K (After Ref.\cite{pourret2011}). }
\end{center}
\end{figure}

A new family of high-temperature superconductors with iron in their composition were discovered in 2008. As far as know, there are only two published studies of the temperature dependence of the Nernst coefficient in this family\cite{pourret2011,kondrat2011}. One of these two has cleraly detected the Nernst signal arising from the thermally-induced motion of superconducting vortices in FeTe$_{0.6}$Se$_{0.4}$\cite{pourret2011}. The data reported is shown in Fig.\ref{Fig_Nernst_vortex_FeSeTe}. As seen in the figure, very much like in the case of cuprates, a sizeable Nernst signal emerges when the resistive transition broadens as a consequence of vortex movement.

According to Eq.\ref{NernstVortex}, one can combine the resistivity and Nernst data to extract the transport entropy associated with each vortex. This quantity, which we call here S$_{vort}$ is simply:

\begin{equation}
\label{NernstVortex}
S_{vort}= \phi_0\frac{N}{\rho}
\end{equation}

Here $\phi_0= 2\times 10^{-15}$ Wb is the magnetic flux of each vortex,  $N=\frac{E_{y}}{\nabla_{x}T}$ is the Nernst signal and $\rho$ the resistivity.  In three dimensions, S$_{vort}$ would be equal to the entropy of each vortex per unit length along its axis. Huebener and collaborators have used this procedure to quantify it in two cuprate superconductors, reporting $6.5\times10^{-15}$ JKm$^{-1}$ in  YBa$_{2}$Cu$_{3}$O$_{7-\delta}$ and $3.7\times10^{-15}$JKm$^{-1}$ in  Bi$_{2}$Sr$_{2}$CaCu$_{2}$0$_{8+x}$ in samples near their optimal doping levels\cite{huebener1995}. Such a quantitative analysis is yet to be done in other families of superconductors.

\section{Review of experiments IV: Quantum oscillations in strong magnetic field}
Magnetic field truncates a three-dimensional Fermi surface to concentric Landau tubes. Electronic orbits become quantized in the plane perpendicular to the magnetic field. Each Landau tube is a one-dimensional Fermi sea of degenerate states.  As the magnetic field is swept, the Landau tubes grow in diameter and exit the Fermi surface one after the other. This leads to quantum oscillations of various measurable physical properties of the system. Oscillations in the field dependence of resistivity and magnetization  are called Shubnikov- de Haas and de Haas-van Alphen effects, after the name of their discoverers. The period of these oscillations is proportional to the extremal section of the Fermi surface perpendicular to the magnetic field. They are detectable only if the temperature and disorder do not smear quantization. This means that the experiment is to be done at low enough temperature (to insure $\hbar \omega_{c} > k_{B}T$) and high enough magnetic field (to satisfy $\omega_{c}\tau > 1$). The temperature dependence of the amplitude of oscillations quantifies the ratio of cyclotron to thermal energies and, therefore,  yields the magnitude of the cyclotron mass\cite{shoenberg1984}.

In a two-dimensional electron gas, the oscillations of the Hall response become quantized in $h/e^2$, leading to the quantum Hall effect. The fundamental reason behind the crucial role of dimensionality resides in the fact that two-dimensional electrons have no kinetic energy along the orientation of magnetic field. Each time the Fermi level lies between two Landau levels, the system is an insulator with no mobile carriers except at its edges.

To reach the quantum Hall regime, one needs to attain a magnetic field strong enough to confine all carriers to a few remaining lowest Landau levels. What happens to a three-dimensional system  in a similar situation is much less explored. For this to happen, the cyclotron energy, $\hbar \omega_{c}$ is to become comparable to the Fermi Energy, $\epsilon_F$. This so-called quantum limit is attained when the Fermi wave-length of electrons become comparable to the magnetic length (See Table 2). For ordinary bulk metals, say copper, the magnetic field necessary to attain this so-called quantum limit is several thousands of Tesla, well beyond the limits of current technology. A field of 10 T corresponds to a magnetic length of $\ell_{B}=(\frac{\hbar}{e B})^{1/2}\sim 8~$nm, an order of magnitude longer than the typical interatomic distance in solids. With a magnetic field of such amplitude, one can reach the quantum limit in metals with a very dilute concentration of mobile electrons. This is the case of stoichiometric semi-metals such as  bismuth and graphite or doped metallic semiconductors. In bismuth, the carrier density of holes is 3$\times$ 10$^{17}$ cm$^{-3}$, which means that there is roughly one itinerant electron per 10$^5$ atoms. Carrier density in graphite is an order of magnitude larger, but it is a layered material with an elongated Fermi surface and a cross-section perpendicular to the high-symmetry axis  as small as bismuth. In both systems, a field of 10 T suffices to attain the quantum limit.

\begin{table}
  \centering
  \begin{tabular}{|c|c|c|}
    \hline
   &First threshold (Quantization) & Second threshold (Quantum Limit)\\
\hline
Time-scale criterion  & $\omega_{c}>\tau^{-1} $ & $\hbar \omega_{c}>\epsilon_{F}$\\
\hline
Length-scale criterion  & $\ell_{B}<(\lambda_{F}\ell_{e})^{1/2}$ &  $\ell_{B} < \lambda_{F}$\\
\hline
 copper  &$\sim 5 T$ & $\sim 5\times10^{4}T$ \\
  \hline
bismuth &$\sim 0.1 T$ & 9 T \\
  \hline
 graphite & $\sim 0.1 T$ & 7.5 T\\
\hline
YBa$_{2}$Cu$_{3}$O$_{7-\delta}$ (p=0.12) & 25 T & $\sim$ 600 T \\
\hline
 \end{tabular}
 \linebreak
\caption{Two distinct limits: Beyond the first threshold field, quantum oscillations become detectable. When the magnetic field exceeds the second threshold filed, electrons are confined to their lowest Landau limit. In semi-metals, such as bismuth and graphite, this limit is accessible with available magnetic fields.}
\label{tableQL}
\end{table}

\begin{figure}
\begin{center}
\resizebox{!}{0.6\textwidth}{\includegraphics{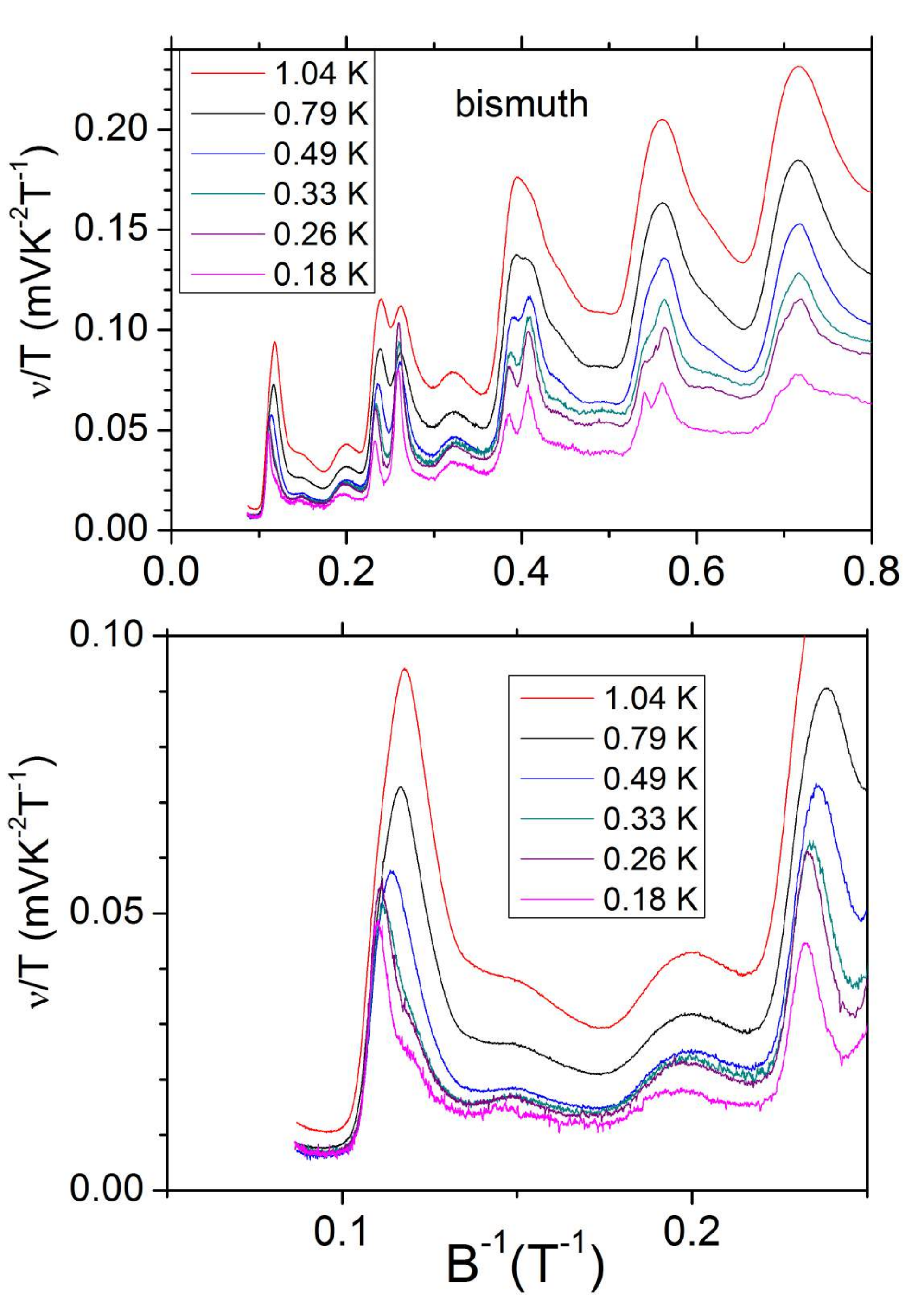}}
\caption{\label{Fig_Bi_Osc} The Nernst coefficient divided by temperature, $\nu/T$, as a function of inverse of magnetic field in bismuth. Quantum oscillations  dominate the non-oscillating background. Warming leads to a substantial increase in $\nu/T$, indicating a substantial contribution by phonon drag (expected to follow a temperature dependence such as T$^{\alpha}$ with $\alpha >1$ ).}
\end{center}
\end{figure}

\begin{figure}
\begin{center}
\resizebox{!}{0.5\textwidth}{\includegraphics{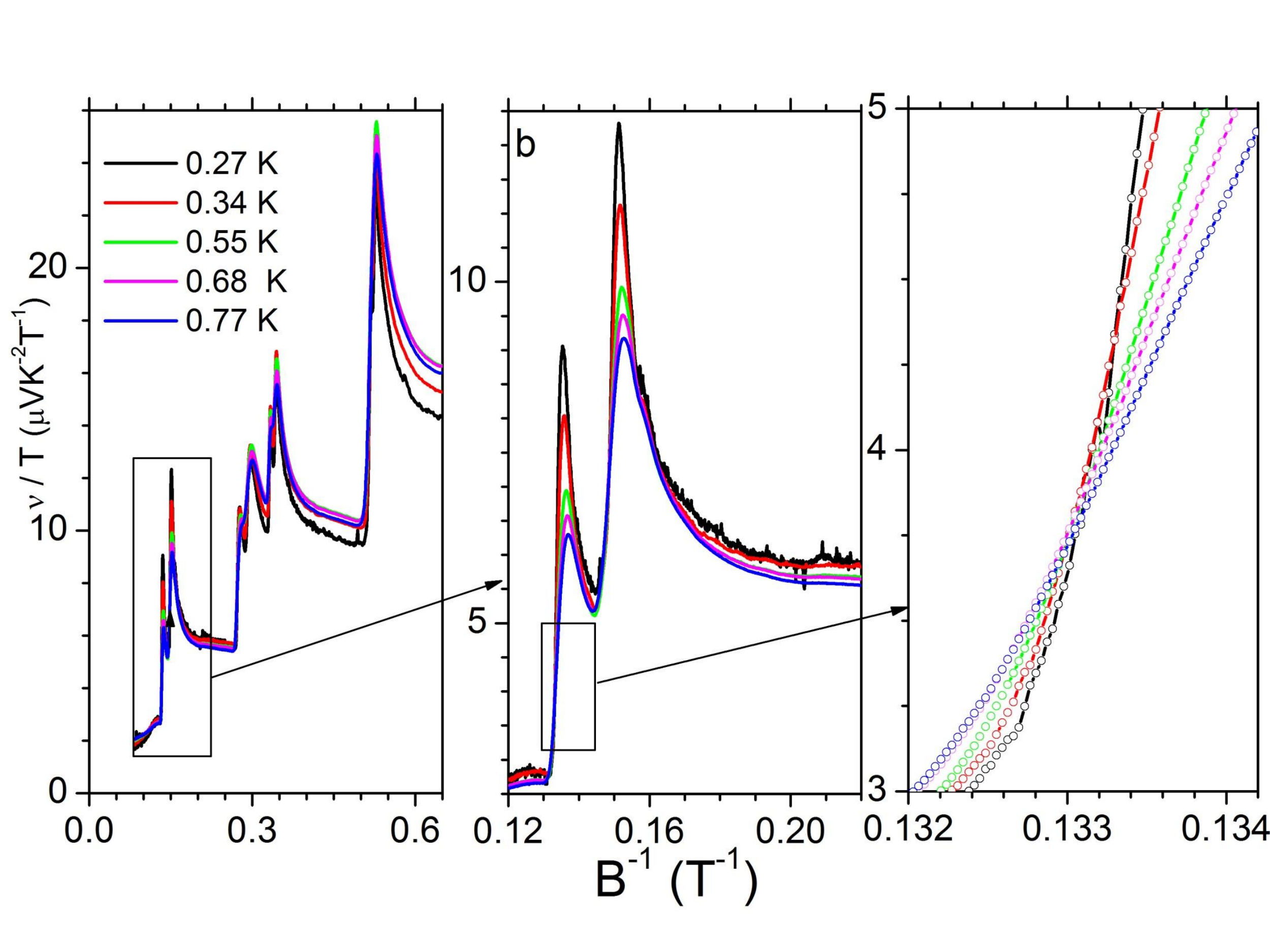}}
\caption{\label{Fig_graphite}The Nernst coefficient divided by temperature, $\nu/T$, as a function of inverse of magnetic field in graphite. Warming does not change the magnitude of $\nu/T$ pointing to the dominance of diffusive component. The middle and the right panels are zooms to a portion of their neighboring panel at their left side. In the right panel, a temperature-independent crossing point becomes visible. Note the sharp asymmetric peaks as in the case of bismuth. }
\end{center}
\end{figure}

During the last few years, unexpectedly large oscillations of the Nernst response was reported in both bismuth\cite{behnia2007b} and graphite\cite{zhu2010,fauque2011} in the vicinity of the quantum limit. In both these systems, when a few Landau tubes remain, a large oscillatory Nernst response dominates the monotonous background. In bismuth, the Nernst response has a large phonon-drag component down to sub-Kelvin temperature range. In graphite, on the other hand, it is mostly diffusive. Nevertheless, the profile of oscillations are qualitatively similar (See Fig.\ref{Fig_Bi_Osc} and \ref{Fig_graphite}).

Early studies both on bismuth\cite{mangez1976} and graphite\cite{woollam1971} had already reported on quantum oscillations of thermoelectric coefficients, but were not extended to low enough temperatures or large enough magnetic fields to detect the spectacular peaks seen when the magnetic field becomes a few tesla and the temperature goes below 1K. More recently, one has detected  giant Nernst oscillations in high-mobility semiconductors such as Bi$_{2}$Se$_{3}$\cite{fauque2013}, SrTiO$_{3-\delta}$ \cite{lin2013} and two systems believed to host massive Dirac electrons, Pb$_{1-x}$Sn$_{x}$Se\cite{liang2013} and BiTeI\cite{ideue2015}. If stoichiometric, these systems are expected to be gapped semiconductors. However, the real samples are sufficiently doped to be pushed to the metallic side of metal-insulator transition and possess a sharp tiny Fermi surface giving rise to a large oscillatory thermoelectric response. Even if it was not known in 2007, giant quantum oscillations in the vicinity of the quantum limit had been already observed decades before. One can find reports as early as 1959 on oscillations of the Nernst coefficient with a large amplitude and low frequency in metals such as elemental zinc\cite{bergeron1959} or aluminium\cite{fletcher1983}. A similar observation was reported in the case of a degenerate semiconductor such as Fe-doped HgSe\cite{tieke1996}.

In two-dimensional systems, following the discovery of the Quantum Hall effect, several experiments on thermoelectric response were performed (See ref. \cite{fletcher1999} for a review).  Thermoelectric coefficients in the Quantum Hall regime have been studied both in  silicon MOSFETs\cite{fletcher1998} and in GaAs/Ga$_{1-x}$Al$_{x}$As hetero-junctions\cite{obloh1986,ying1994,tieke1997,tieke1998}. More recently, graphene has attracted tremendous attention as a two-dimensional electron gas displaying Quantum Hall effect. Its thermoelectric response in the quantum Hall system was subject a few recent studies\cite{zuev2009,wei2009,checkelsky2009}. In contrast to the case of heterojunctions where the phonon drag contribution dominates the thermoelectric response down to sub-Kelvin temperature\cite{ying1994,tieke1998}, the thermoelectric response in graphene is essentially diffusive and even at 10 K, phonon drag is negligible.

\begin{figure}
\begin{center}
\resizebox{!}{0.7\textwidth}{\includegraphics{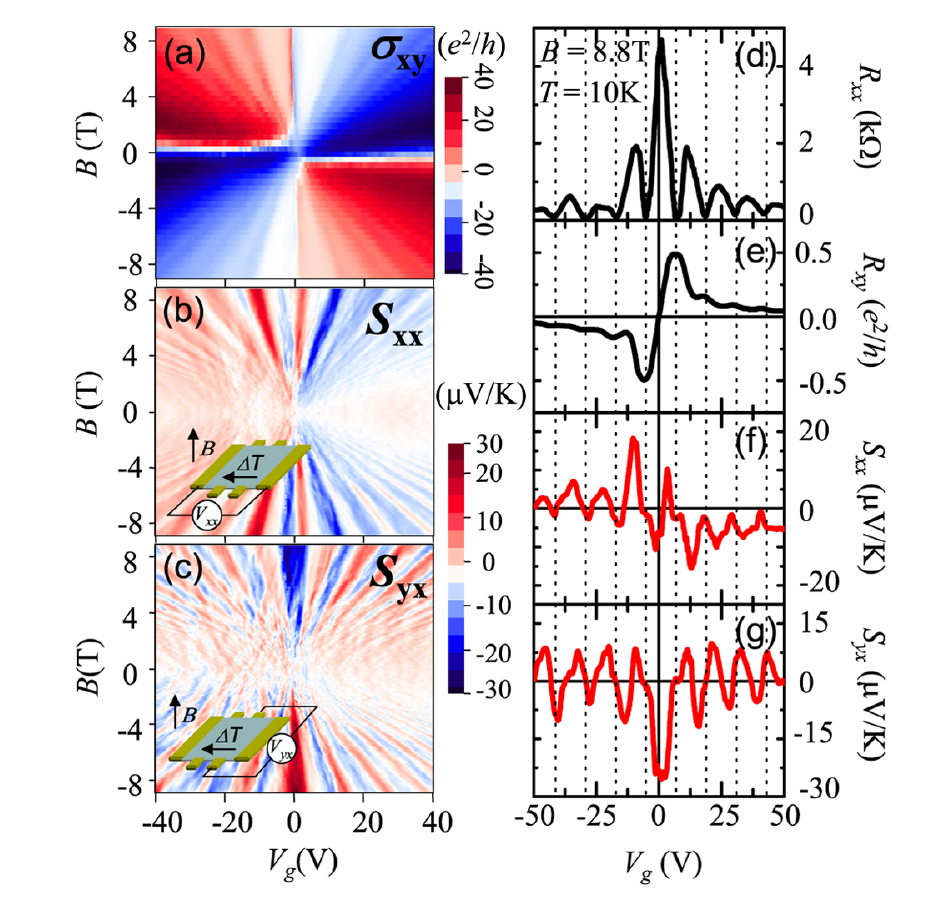}}
\caption{\label{Fig_graphene}Transport coefficients in graphene as a function of gate voltage in a fixed magnetic field [After Zuev \emph{et al.} 2009]. Sweeping the chemical potential across the Landau levels  induces anomalies in all transport coefficients. The Nernst coefficients presents two peaks of opposite signs at each passage between successive Landau levels [After Zuev \emph{et al.}\cite{zuev2009}]. }
\end{center}
\end{figure}

\begin{figure}
\begin{center}
\resizebox{!}{0.6\textwidth}{\includegraphics{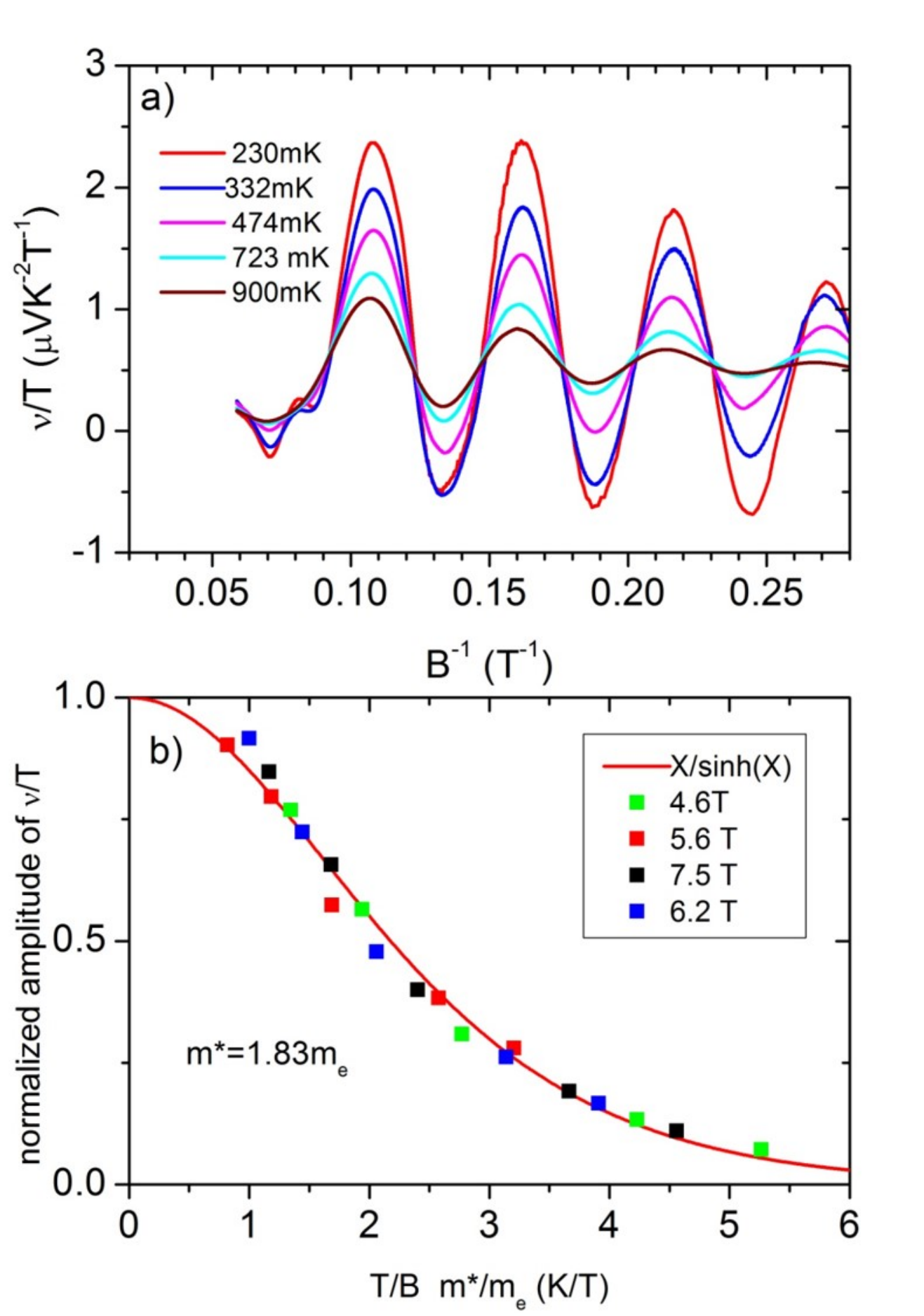}}
\caption{\label{Fig_STO_Osc} Top: The Nernst coefficient divided by temperature, $\nu/T$, as a function of inverse of magnetic field in SrTiO$_{3}$\cite{lin2013}. The profile of oscillations are qualitatively different from those seen in bismuth and graphite and do not present asymmetric sharp peaks. Bottom: The temperature dependence of the amplitude of oscillations is in agreement with what is expected in the Lifshitz-Kosevitch formalism.}
\end{center}
\end{figure}

The thermoelectric response in a two-dimensional electron gas subject to a strong magnetic field has been the subject of theoretical investigations dating back as early as 1984\cite{jonson1984,oji1984,cooper1997}. The three-dimensional case and its Nernst quantum oscillations were subject of a number of recent studies\cite{bergman2010,lukyanchuk2011,sharlai2011}.

A number of empirical observations have been reported on the profile of the quantum oscillations of the Nernst response. Studying a two-electron gas in GaAs/Ga$_{1-x}$Al$_{x}$As heterostructures, Tieke \emph{et al.}\cite{tieke1997} found that the Nernst response is proportional to the field derivative of the Seebeck coefficient. In other words, the transverse response presents peaks which are shifted by $\pi/2$  compared to those seen in the longitudinal response. They provided a justification for this observation based on the Mott formula. Such a shift can also be seen in the case of graphene(See fig.\ref{Fig_graphene}). As seen in the figure,  each Nernst oscillation consists of a pair of negative and positive peaks sandwiching a vanishing signal. This is in qualitative agreement with the theoretical expectations\cite{jonson1984}. The Nernst response is expected to vanish twice, at the center of a Hall plateau and at the passage from one plateau to the next. In both cases, the Hall mobility presents an extremum as a function of chemical potential.   Nernst quantum oscillations in graphite, on the other hand, present a very different profile. In this case, each oscillation has an asymmetric sharp peak (See Fig. \ref{Fig_graphite}). This empirical observation suggested a qualitative change in the transverse thermoelectric response in the passage from 2D to 3D\cite{zhu2010}.

In a theoretical work devoted to the field dependence of $\alpha_{xy}$ (dubbed ``dissipationless  Nernst effect''), Bergman and Oganesyan\cite{bergman2010} derived the following expression for two dimensions:

 \begin{equation}
\label{Berg_Oga2D}
\alpha_{xy}^{2D}=\frac{ek_{B}}{h}\sum_{n=0}^{n_{max}}[f_{n}\ln f_{n}+(1-f_{n})\ln (1-f_{n})]
\end{equation}

Here, $f_{n}=f^{0}$ is the Fermi distribution of electrons belonging to the Landau level indexed n. The field dependence of $\alpha_{xy}^{2D}$ consists of sign-changing oscillations with an amplitude set by the quantum of thermoelectric conductance. In three dimensions, they found another expression:

 \begin{equation}
\label{Berg_Oga3D}
\alpha_{xy}^{3D}=-\frac{ek_{B}}{h}\frac{\pi^{2}}{3}\sum_{n=0}^{n_{max}}\frac{k_{B}T}{h v_{Fn}}
\end{equation}

In this equation, $v_{Fn}$ is the Fermi velocity of the n-th Landau tube. According to this latter equation, the three-dimensional $\alpha_{xy}$ should present asymmetric sharp peaks at each evacuation of a Landau level, leading to a drastic decrease in $v_{Fn}$. This theoretical prediction  was in agreement with the experimental observation of the field-dependence of the Nernst coefficient $\nu$\cite{zhu2010}. When the magnetoresistance is field-linear, $\nu$ is simply proportional to $\alpha_{xy}$. Moreover, experiment resolved temperature-independent crossing points\cite{fauque2011}, reminiscent of those predicted by theory\cite{bergman2010} (See Fig. \ref{Fig_graphite}).

Later studies on other three-dimensional solids brought more information and additional complications. Giant oscillations of the Nernst coefficient were indeed observed in doped semiconductors such as Bi$_{2}$Se$_{3}$\cite{fauque2013} and SrTiO$_{3-\delta}$ \cite{lin2013}. However, as one can see in Fig. \ref{Fig_STO_Osc}, these oscillations do not consist of asymmetric peaks. As in the case of dHvA and SdH effect, the temperature dependence of the amplitude of oscillations could be used to extract the effective mass of carriers using the Lifshitz-Kosevitch formalism.  Thus, as far as the experiment can tell, there is a clear qualitative difference between the profile of quantum oscillations of the Nernst coefficient in compensated semi-metals such as bismuth and graphite and in doped semiconductors.

To explain this difference, two distinct possibilities can be thought of.  First, in compensated metals, the Hall resistivity remains much smaller  than longitudinal resistivity. Therefore, since magnetoresistance is field-linear, $\nu$, remains within a good approximation proportional to $\alpha_{xy}$. This is not the case of uncompensated systems in which the field dependence of $\alpha_{xy}$ and $\nu$ could be very different. The expected singularity in $\alpha_{xy}$ may not be directly visible in  uncompensated dilute metals. Another road to an explanation may reside in the fact the carrier concentration in compensated semimetals does not remain constant in the vicinity of the quantum limit. A \emph{quantitative} explanation of the sharp asymmetric peaks seen in the magnetostriction of bismuth\cite{kuechler2014} invokes this field-induced variation in carrier concentration, which is absent in uncompensated doped semiconductors. It may be that the additional carriers brought to the system to preserve charge neutrality play a role in generating the sharp Nernst peaks each time the height of a Landau tube\cite{shoenberg1988} is reduced to zero. Future investigations and in particular a direct comparison of the profile of $\alpha_{xy}$ (and not $\nu$) in different dilute metals would be instructive.

In spite of the absence of a quantitative understanding of the Nernst response in the vicinity of the quantum limit in different bulk metals, the sensitivity of the Nernst effect makes it a very useful probe of Landau spectrum. In the case of bismuth, angle-resolved Nernst experiments  has mapped the complex Landau spectrum for the whole solid angle up to 28 T. This complex spectrum has been found to be in satisfactory agreement with theoretical calculations, pinning down the band parameters of bismuth with a high precision\cite{zhu2011,zhu2012}. Moreover, the angle-resolved study  led to the identification of the origin of the unexpected Nernst peaks observed in an initial experiment\cite{behnia2007c}. These peaks were due to presence of minority domains in a twinned bismuth crystal. This was revealed by comparing their evolution with rotation with those expected by theory\cite{zhu2012} for a crystal rotated by 108 degrees, the  angle imposed by twinning in bismuth\cite{hall1954}.

\section{Other Nernst experiments}
\subsection{The puzzling Nernst coefficient of URu$_{2}$Si$_{2}$} A very recent report on the Nernst coefficient in this heavy-fermion system points to a puzzle, which has yet to be figured out\cite{yamashita2015}. URu$_{2}$Si$_{2}$ is host to a phase transition occurring at 18 K and leading to the emergence of the so-called hidden order\cite{mydosh2011}. In 2004, it was reported that the entry to the hidden order is accompanied by a sudden enhancement in the Nernst response\cite{bel2004}. This can be understood by taking into account the sudden decrease in the carrier concentration and the drastic increase in the carrier mobility in the hidden order. Both of them contribute to amplify the Nernst coefficient\cite{behnia2009}. One can destroy the hidden order by the application of a strong magnetic field. This destruction is concomitant with the suppression of the large Nernst response\cite{levallois2009}.

\begin{figure}
\begin{center}
\resizebox{!}{0.7\textwidth}{\includegraphics{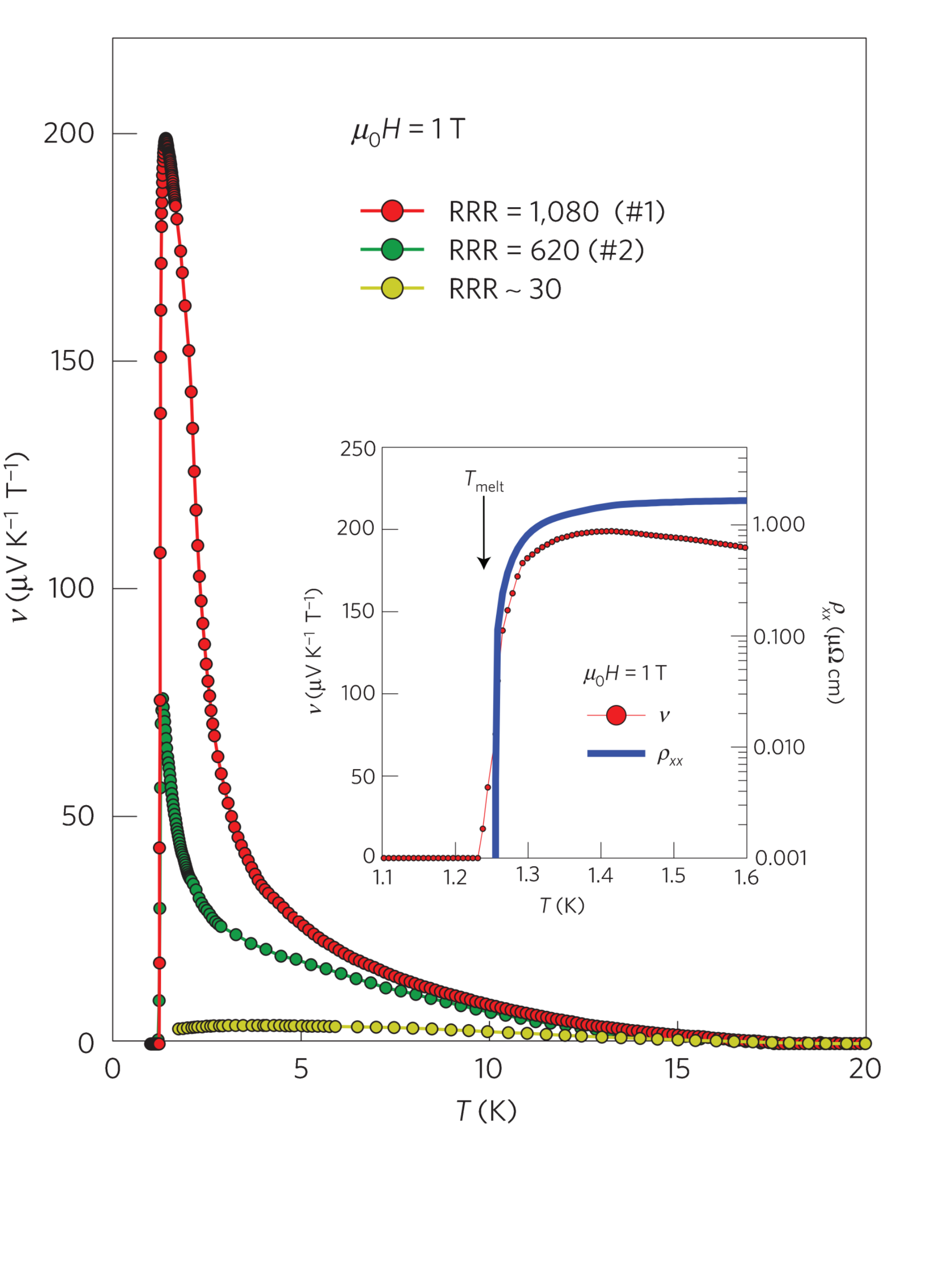}}
\caption{\label{Fig_URUSI} Main panel: The temperature dependence of the Nernst coefficient, $\nu$, at 1 T in crystals of URu$_{2}$Si$_{2}$ with different RRR\cite{yamashita2015}. Anomalously large Nernst signal and thermomagnetic figure of merit. Inset: The Temperature dependence of $\nu$, (left scale) and resistivity (right scale) measured at 1 T near the superconducting transition. The Nernst signal in the normal state is by far larger than what is measured in the vortex liquid state. Reprinted by permission from Macmillan Publishers Ltd\cite{yamashita2015}.}
\end{center}
\end{figure}

 These experiments were performed in samples with a RRR of the order of 30 (or a residual resistivity of about 10 m$\Omega$cm).  Yamashita \emph{et al.} recently studied new generation samples of URu$_{2}$Si$_{2}$ which were  much cleaner (their RRR was 620 and 1080). They found that in these samples, the enhancement of the Nernst coefficient triggered by the 18 K phase transition is even more pronounced\cite{yamashita2015} compared to the older generation\cite{bel2004}(See Fig.\ref{Fig_URUSI}). The low-temperature mobility is much higher in these new-generation crystals and qualitatively, the enhancement of the Nernst coefficient with carrier mobility is conform to Eq.~\ref{nuqp}. However, Yamashita \emph{et al.} observed also another enhancement of the Nernst coefficient at T$^{*}\simeq 5$ K (i.e. above the superconducting transition temperature  (T$_{c}\simeq 1.5$ K). This second enhancement was absent in the data on dirtier crystals\cite{bel2004}. They attributed this  to superconducting fluctuations\cite{yamashita2015} and, noticing that the magnitude of the observed signal is much larger than what is expected for gaussian fluctuations, speculated about an exotic source of superconducting fluctuations. What is currently known about various sources of a Nernst signal inspires us a couple of comments.

First of all, one is inclined to compare  URu$_{2}$Si$_{2}$ with other superconductors, in which one has detected both vortex (below T$_{c}$) and fluctuating (above T$_{c}$) contributions to the Nernst response. In all of them, the fluctuating signal in the normal state is a high-temperature tail of the vortex signal. In the case of URu$_{2}$Si$_{2}$,  on the other hand no vortex Nernst signal has been clearly resolved. The contrast is visible by comparing the inset of Fig. \ref{Fig_URUSI} with Fig. \ref{Fig_Nernst_vortex_YBCO} and Fig. \ref{Fig_Nernst_vortex_FeSeTe}. If superconductivity generates the large Nernst signal detected above T$_c$, then one would have expected a much larger Nernst signal below T$_{c}$. This is not the case. Note that experiment has detected a first-order transition identified as the melting transition of the vortex lattice\cite{okazaki2008}. Therefore, the temperature range for the expected emergence of the vortex Nernst signal is known. Since there is no enhanced Nernst signal in this range, any Nernst signal by mobile vortices should be be much lower in amplitude than the signal observed by experiment above T$_{c}$.

Alternatively, to check a possible quasi-particle origin of the observed signal, it is instructive to compare the main panel of Fig. \ref{Fig_URUSI} with the case of bismuth (See Fig.\ref{Fig_Bi}). The Nernst coefficient even in the cleanest URu$_{2}$Si$_{2}$ remains well below what has been observed in the dirtiest bismuth. Moreover, in both cases, the enhancement of RRR leads to an amplification of the Nernst coefficient.

In section 3, we argued the magnitude of $\alpha_{xy}$ for a given type of carriers is set by the ratio of entropy they transport to the magnetic flux they carry. Thus, quasi-particles with a long mean-free-path and a long Fermi wave-length can generate a large $\alpha_{xy}$ exceeding by far what can be produced by either vortices or short-lived Cooper pairs. The $\alpha_{xy}$ in URu$_{2}$Si$_{2}$  peaking to 23 AK$^{-1}$cm$^{-1}$ (at T$\simeq$1.65 K and B$\simeq$1 T)\cite{yamashita2015} is indeed ``colossal'' compared to the highest value seen in a superconductor such as optimally-doped YBa$_{2}$Cu$_{3}$O$_{7-\delta}$, where $\alpha_{xy}$ peaks to 0.08 AK$^{-1}$cm$^{-1}$ (at T$\simeq$80 K and B$\simeq$ 12 T\cite{ri1994}), but remains smaller than what has been seen in a bismuth crystal even a RRR as low as 65 ($\alpha_{xy}\sim$ 120 AK$^{-1}$cm$^{-1}$ at T$\simeq$1 K and B$\simeq$ 0.1 T\cite{behnia2007}).

These comparisons suggest that superconducting fluctuations are an unlikely source of a Nernst signal as large the one observed in the normal state of URu$_{2}$Si$_{2}$. On the other hand, a scenario invoking quasi-particles needs to explain what happens at T$^{*}$ to the Fermi surface (or more likely to one of its components) in order to produce the enhanced Nernst signal. There is no satisfactory answer to this question. However, a similar mystery exists in the case of PrFe$_{4}$P$_{12}$\cite{pourret2006b}. In this system, ordering at 6 K leads to an enhancement of the Nernst response. At 2~K, well below the ordering temperature, a second jump occurs in the Nernst coefficient, which becomes as large as 60$\mu$ VK$^{-1}$ (comparable to the peak value in URu$_{2}$Si$_{2}$ with RRR=620). In non-superconducting PrFe$_{4}$P$_{12}$, this is undoubtedly caused by a sudden jump in quasi-particle mobility, suggesting that such a scenario cannot be ruled out in the case of URu$_{2}$Si$_{2}$.

\subsection{Anomalous Nernst effect and the Berry phase}
Let us briefly mention Anomalous Nernst Effect (ANE), which has attracted significant attention in the past few years. This refers to the thermoelectric counterpart of the Anomalous Hall Effect (AHE). In ferromagnetic solids, there is a spontaneous Hall current on top of the normal Hall effect. Recent work has shown that this can be traced back to the presence of an anomalous velocity intimately linked with Berry-phase curvatures(See Nagaosa \emph{et al.}\cite{nagaosa2010} for a review).

Following its introduction by Berry in 1984\cite{berry1984}, the idea that tuning the geometric phase of a quantum state can lead to experimentally observable physical phenomena has deeply influenced the understanding of diverse phenomena in condensed-matter physics\cite{xiao2010}. In the case of Bloch waves in crystals, the Berry curvature modifies the group velocity by introducing an anomalous term \cite{xiao2006}:
\begin{equation}
\label{Ano_Velo}
\dot{r}=\frac{1}{\hbar}\frac{\partial\epsilon_{n}(\bf{k})}{\partial \bf{k}}+\frac{e}{\hbar}\bf{E}\times\Omega_{n} (\bf{k})
\end{equation}

In the absence of the second term in the right side of this equation, the group velocity, $\dot{r}$ is set by the energy dispersion, $\epsilon_{n}(\bf{k})$ of a band indexed $n$. Various transport properties of electrons in a crystal would be affected by the presence of a finite $\Omega_{n} (\bf{k})$. The latter is called Berry curvature. It can be defined for the Bloch wave-function of a band indexed $n$:

\begin{equation}
\label{Bloch}
\Psi_{\bf{k}n}(r)=exp(n\bf{k}\bf{r})u_{\bf{k}n}(\bf{r})
\end{equation}

in the following way\cite{mikitik1999}:
\begin{equation}
\label{Berry-curv}
\Omega_{n} (\bf{k})=\int u_{\bf{k}n}^{*}(\bf{r})\nabla_{k}u_{\bf{k}n}(\bf{r})
\end{equation}

When the $\Omega_{n} (\bf{k})$ field is such that its integral along a closed orbit remains finite one speaks of s non-trivial Berry phase. This is  a generalization of the physics of Aharanov-Bohm effect to a situation in which the phase difference is nor produced by a \emph{real} magnetic field, but by an \emph{effective} magnetic field described by the Berry curvature.

In a theoretical study of ANE, Xiao and co-workers\cite{xiao2006} argued that in spite of the fact that the force produced by a thermal gradient is statistical and therefore intrinsically macroscopic (in contrast to the one generated by an electric field), a finite Berry curvature generates an anomalous transverse thermoelectric response, $\alpha^{A}_{xy}$. They found that the  magnitude of the latter is linked to the the anomalous Hall effect, $\sigma^{A}_{xy}$, :
 \begin{equation}
\label{Mott-ANE}
\alpha^{A}_{xy}=\frac{\pi^{2}}{3}\frac{k_{B}}{e}k_{B}T\frac{\partial\sigma^{A}_{xy}}{\partial\epsilon}|_{\epsilon=\epsilon_{F}}
\end{equation}
This is the Mott formula, the same which links the normal versions of the two effects in the purely orbital case. A number of experimental studies have been devoted to the ANE in ferromagnetic solids\cite{lee2004,pu2008,ramos2014}. In two systems, the ferromagnetic semiconductor, Ga$_{1−x}$Mn$_{x}$As\cite{pu2008} and  half-metallic Magnetite(Fe$_{3}$O$_{4}$)\cite{ramos2014}, the measured values of ANE and AHE were reported to be in agreement with the Mott formula.

One expects to see in near future many other Nernst experiments probing electronic states with non-trivial topology.

\section{Concluding remarks}
Contrary to what one can still read here and there in scientific literature, the Nernst effect is not negligibly small in Fermi liquids.  Quasi-particles generate a Nernst signal which can vary by many orders of magnitude in different Fermi liquids. The largest known Nernst coefficients are recorded in dilute metals hosting extremely mobile electrons. On the other hand, when the quasi-particle contribution is tuned to zero in the normal-state of a dirty superconductor, one can resolve a sizeable Nernst signal coming from fluctuating Cooper pairs. Mobile superconducting vortices are another well-established source of a Nernst signal below the critical temperature. Finally, there are indications that squeezing a Landau tube generates Nernst peaks, which become visible in the vicinity of quantum limit.

This work was supported by the Agence Nationale de la Recherche through the SUPERFIELD project. We thank Beno\^{\i}t Fauqu\'e, Francis Lalibert\'e, Xiao Lin, Alexandre Pourret, Louis Taillefer Cheng-Long Zhang and Zengwei Zhu for discussions and comments.

\end{document}